\documentclass[preprint,amsmath,amssymb,aip,jcp]{revtex4-2}
\usepackage{graphicx}
\usepackage{bm}
\usepackage{xcolor}
\usepackage{ragged2e}
\usepackage{epstopdf}
\usepackage{booktabs}
\usepackage{multirow}
\usepackage[colorlinks=true,linkcolor=black,citecolor=blue,urlcolor=blue]{hyperref}
\newcommand{\be}{\begin{equation}}
\newcommand{\ee}{\end{equation}}
\newcommand{\bea}{\begin{eqnarray}}
\newcommand{\eea}{\end{eqnarray}}

\newcommand{\bee}{\begin{eqnarray*}}
\newcommand{\eee}{\end{eqnarray*}}

\begin{document}
\title{Electrohydrodynamic lubrication theory for an immersed cylinder moving near wall}
\author{Anirban Chatterjee}
\affiliation{Univ. Bordeaux, CNRS, LOMA, UMR 5798, 33405 Talence, France.}
\author{Yacine Amarouchene}
\affiliation{Univ. Bordeaux, CNRS, LOMA, UMR 5798, 33405 Talence, France.}
\author{Thomas Salez}
\thanks{thomas.salez@cnrs.fr}
\affiliation{Univ. Bordeaux, CNRS, LOMA, UMR 5798, 33405 Talence, France.}
\date{\today}
\begin{abstract}
The free motion of charged colloids within ionic solutions and in the vicinity of charged boundaries, is a phenomenon that occurs in various natural, biological and industrial settings. Here, we develop an electrohydrodynamic lubrication theoretical framework, in order to characterize such a motion in the case of an infinite rigid cylinder near a rigid wall. Combining hydrodynamic lubrication theory, Debye-H\"uckel electrostatics, and Nernst-Planck electrokinetics, we derive the three coupled equations of motion for the normal, longitudinal and rotational degrees of freedom of the cylinder, which are then investigated numerically and through asymptotic analysis. Our results reveal complex behaviours, beyond existing asymptotic electroviscous-lift expressions, and extend the classical Faxen-Brenner-like mobility matrix when surface charges and dissolved ions are incorporated.
\end{abstract}
\maketitle

\section{Introduction}
The motion of a rigid particle immersed in a viscous fluid close to a wall 
is a fundamental problem in low-Reynolds-number hydrodynamics, with far-reaching implications for colloidal science, micro/nanofluidics, tribology, and biological transport~\cite{Leal2007,HappelBrenner1983,bocquet2010nanofluidics}.
In the case of fore-aft symmetric objects in the time-reversal-symmetric steady-Stokes regime, such a two-fold symmetry forbids 
the generation of lift forces normal to the direction of motion, unless an additional 
physical ingredient is added in order to break the fore-aft symmetry~\cite{Skotheim2005,Bureau2023,Rallabandi2024}. Over the past decades, several symmetry-breaking mechanisms have been identified and studied. Examples include wall or particle elasticity~\cite{Saintyves2016, davies2018elastohydrodynamic, rallabandi2018membrane, vialar2019compliant, zhang2020direct, fares2024observation}, 
surface slippage inhomogeneities~\cite{Karan2020, rinehart2020lift}, viscoelasticity~\cite{Pandey2016,Bharti2024,oratis2025viscoelastic}, odd-viscosity fluids \cite{lou2022odd}, 
and electrokinetic effects which are the focus of the present work. 

From an applied standpoint, the interplay between hydrodynamic and electrokinetic 
forces near charged walls is central to a number of technologies. For instance, in hydrodynamic 
chromatography and field-flow fractionation, the equilibrium position 
of colloidal particles in a channel determines the elution time and hence the 
separation selectivity. An electroviscous lift should directly affect this 
position~\cite{Small1974, Giddings1993}. In microfluidics, the transport, focusing, 
and sorting of cells, vesicles, and nanoparticles in electrolyte-filled channels 
routinely involves the very configuration at stake here~\cite{Li2014,Xuan2019}. 
At the nanoscale, water lubrication of ceramic tribological pairs produces 
extremely thin films between charged surfaces, where electroviscous effects 
significantly modify the load-bearing capacity~\cite{ZhangUmehara1998,WongETal2003}.

When two neighbouring charged solids are in relative motion in an electrolyte, the
flow within the thin gap between them convects the mobile ions of the electric double layers (EDLs) 
that form at solid-liquid interfaces, thereby generating a streaming current. In the absence of an 
external circuit, this current must be balanced by a conduction current, which requires 
the establishment of a streaming-potential gradient along the gap. The resulting 
electric field acts back on the fluid through the electric force exerted on the charges within 
the EDLs, modifying the effective viscous dissipation. This feedback loop, commonly 
referred to as the electroviscous effect, leads to an apparent increase in viscosity and 
alters the forces and torques experienced by the solids. The phenomenon was first described theoretically by 
Elton~\cite{Elton1948} and has since been studied extensively in channel 
flows~\cite{RiceWhitehead1965,Levine1975,Hunter1981,RusselSavilleSchowalter1989,LiJin2008,ChakrabortyChakraborty2011}, as well as in confined particle-wall geometries~\cite{cramail2024forces}.

In the context of lubrication flows, the electroviscous coupling has a particularly 
striking consequence: it can generate a normal lift force on an immersed charged particle translating 
parallel to a charged wall, even in the creeping-flow regime where inertial lift 
is absent~\cite{Bureau2023}. This electroviscous lift arises because the streaming-potential-induced 
backflow modifies the pressure distribution asymmetrically, producing a net normal 
force. The effect was first predicted theoretically by Bike and Prieve~\cite{Bike1990}, 
who analyzed the normal and longitudinal motions of a charged sphere near a charged 
wall in the limit of thin double layers, \textit{i.e.} $\kappa a \gg 1$, where $\kappa^{-1}$ is 
the Debye screening length and $a$ the particle radius. 
Warszy\'nski and van de Ven~\cite{Warszynski2000} subsequently treated the 
electroviscous drag on a charged cylinder moving longitudinally near a charged wall, in the thin-EDL limit and at low P\'eclet (Pe) number, where Pe measures the dimensionless ratio between ionic advection and diffusion. 
Tabatabaei \textit{et al.}~\cite{Tabatabaei2006} extended such a model to include both 
longitudinal and rotational motions of the cylinder, using the Lorentz reciprocal theorem. 
A general theoretical framework for the thin-EDL regime was then developed by 
Yariv, Schnitzer, and Frankel, in a series of works~\cite{Yariv2011,Schnitzer2012,Schnitzer2016} 
covering successively the low-P\'eclet-number and moderate-P\'eclet-number regimes, as well as the high-P\'eclet-number regime where the shear-induced electroviscous repulsion 
becomes dominant.

Beyond the thin-EDL limit, the situation is considerably less explored. 
Liu \textit{et al.}~\cite{Liu2018} studied the normal motion of a sphere in the 
regime where $\kappa a \sim O(1)$ with charge-regulation boundary conditions.
Zhao \textit{et al.}~\cite{Zhao2020} analyzed the normal motion of a 
sphere for three types of boundary conditions (constant charge, constant potential, 
and charge regulation), and Rodr\'iguez Matus \textit{et al.}~\cite{Matus2022} 
obtained the electroviscous drag for the normal motion of a sphere. However, 
as pointed out by Chun and Ladd~\cite{ChunLadd2004}, the Debye--H\"uckel approximation 
itself can introduce substantial discrepancies between constant-charge and 
constant-potential predictions, making it important to investigate the regime 
of moderate screening ($\kappa a \sim O(1)$) without further limiting assumptions.

Despite the above theoretical body of literature, experimental evidence for the electroviscous 
lift has remained limited~\cite{Bureau2023}.
Early indirect observations relied on tracking particle equilibrium heights under 
shear in the presence of gravity~\cite{AlexanderPrieve1986, BikePrieve1995}, or 
on inferring forces through evanescent-wave scattering~\cite{Bike2002}. 
These preliminary measurements underscore the need for direct quantitative experiments, as well as for advanced theoretical models 
that can capture the full richness of the electrohydrodynamic coupling. 

In the present work, we address the normal, longitudinal and rotational motions of an infinite rigid cylinder near a flat rigid wall, in presence of surface charges and dissolved ions, within the lubrication approximation and at moderate P\'eclet numbers. Our approach 
differs from prior studies in two key respects. First, we do not impose the 
thin-EDL condition (\textit{i.e.} $\kappa a \gg 1$). This is achieved by directly solving 
the linearized Poisson-Boltzmann equation within the 
lubrication gap. Second, and most importantly, we address together the normal, 
longitudinal, and rotational degrees of freedom of the moving object, which are coupled by the governing electrohydrodynamic equations, hence allowing us to extend the Faxen-Brenner-like microhydrodynamic mobility matrix to the charged case. The coupled formulation we propose relies on the theoretical framework developed for soft-lubrication problems in our previous works~\cite{SalezMahadevan2015,Bertin2022,Jha2024,Bharti2024}, where a rich variety of 
unexpected inertial-like dynamical behaviors was uncovered. Here, the symmetry-breaking mechanism is of 
electrokinetic origin rather than elastic, but the mathematical structure is analogous.

The resulting equations of motion of the cylinder, which consist in a system of three coupled nonlinear ordinary differential equations, are solved numerically for three canonical configurations: i) normal motion under external normal (gravity-like) forcing; 
ii) normal and longitudinal motions under normal and longitudinal forcing, but with frozen rotation; and iii) normal, longitudinal and rotational motions under normal forcing and with an initial longitudinal speed. From the obtained numerical solutions, we observe the emergence of a generalized electroviscous lift force, beyond classical asymptotic expressions. Moreover, after possible transient oscillations, the longterm steady gap height is eventually set by a balance between normal forcing, electrostatic repulsion, and longitudinal-motion-induced electroviscous lift. 

The remainder of the present article is organized as follows. In Section~II, we present 
the mathematical model, including the hydrodynamic lubrication equations, the 
Debye-H\"uckel electrostatic theory, and the streaming-potential formulation through the Nernst--Planck equation, before 
deriving the resulting coupled equations of motion for the cylinder. In Section~III, we present and discuss the numerical solutions of the equations of motion for the three canonical configurations, along with analytical asymptotes in particular regimes.
Finally, Section~IV provides conclusions and perspectives. 

\section{Model}
\begin{figure}[t!]
\centering
\includegraphics[width=0.7\linewidth]{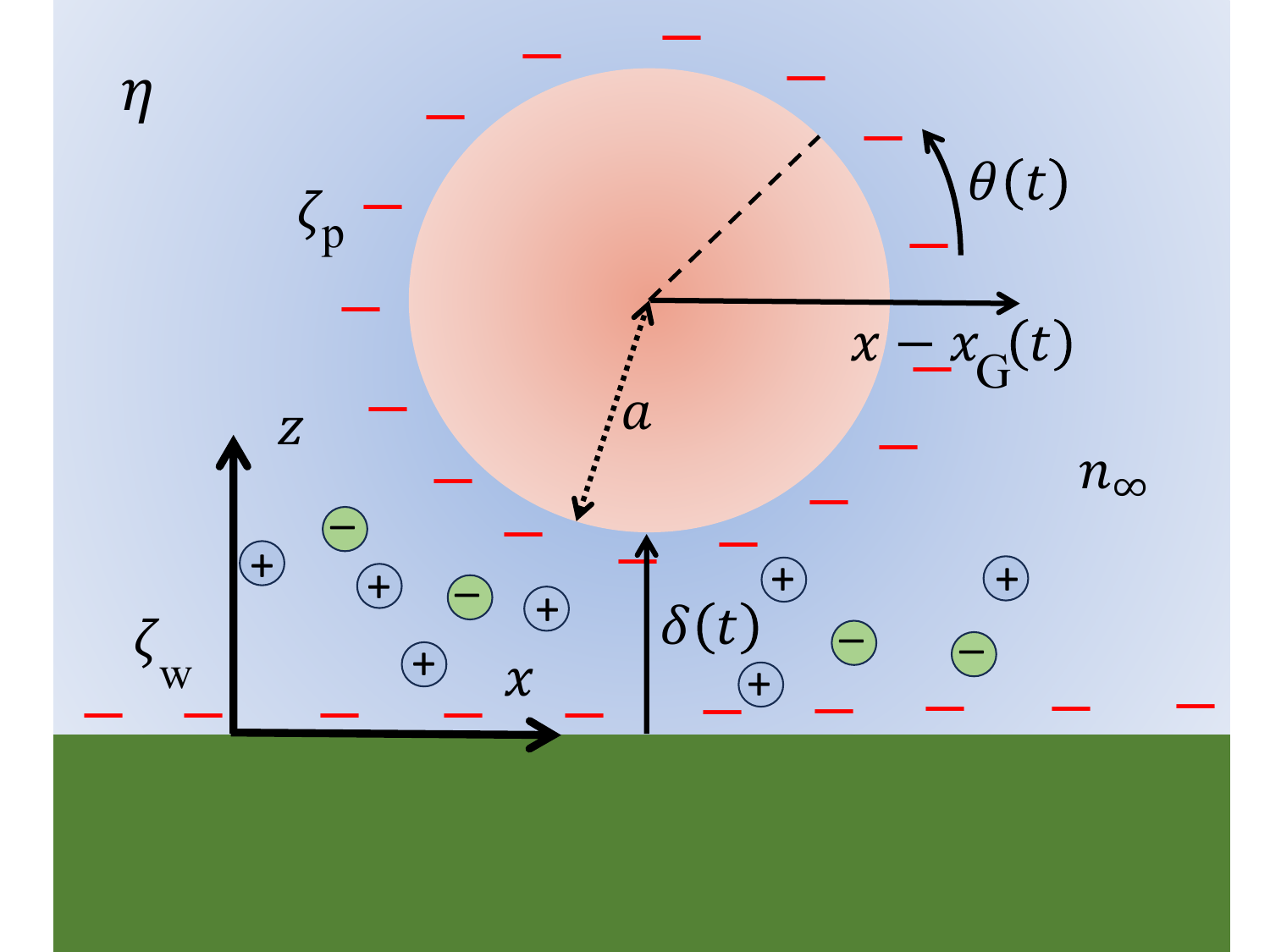}
\caption{\textit{Schematic of the system. An infinite rigid cylinder (pink) of radius $a$ moves in a viscous fluid (blue) of dynamic shear viscosity $\eta$, near and above a rigid flat wall (green), along time $t$. Both the surfaces of the wall and the cylinder are negatively charged and they carry constant surface electric potentials, $\zeta_{\textrm{w}}$ and $\zeta_{\textrm{p}}$ respectively. The fluid is electroneutral but has positive and negative ions dissolved in it, with a bulk concentration $n_{\infty}$ per ionic species. The three degrees of freedom of the cylinder are the normal (along $z$), longitudinal (along $x$) and rotational positions (with respect to the horizontal direction $x$), denoted by $\delta(t)$, $x_{\textrm{G}}(t)$, and~$\theta (t)$, respectively.}}
\label{Schematic1}
\end{figure}
Figure~\ref{Schematic1} presents the system under investigation. We consider an infinite rigid cylinder of radius $a$, density $\rho$, surface electric potential $\zeta_{\textrm{p}}$, within a Newtonian fluid of dynamic shear viscosity $\eta$. We further consider the motion along space $(x,z)$ and time $t$ of the cylinder to always occur in the vicinity of a horizontal flat rigid wall, with surface electric potential $\zeta_{\textrm{w}}$. Moreover, the fluid is assumed to be an electrolyte, containing an overall-neutral assembly of positive and negative ions, with symmetric valencies $z_{+}=\tilde{z}$ and $z_{-}=-\tilde{z}$ respectively, where $\tilde{z}>0$, local concentrations $n_{\pm}(x,z,t)$, and identical bulk concentrations $n_{\infty}$ as well as identical diffusion constants $D$. We consider three degrees of freedom for the motion of the cylinder: the normal distance $\delta(t)$ to the rigid wall along the vertical direction $z$, the longitudinal coordinate $x_\textrm{G}(t)$ of the center of mass along the wall, and the rotational coordinate $\theta(t)$ with respect to the horizontal direction $x$. We initially consider the small-gap limit, \textit{i.e.} $\delta_0 \equiv \delta(t=0) = \epsilon a$, where we introduced the lubrication parameter $\epsilon$ satisfying $\epsilon \ll 1$, and we assume this scale separation to hold at all times.

In this limit, the surface profile of the cylinder can be approximated by a parabola, so that the thickness profile of the fluid film reads :
\begin{equation}
h(x,t) \simeq \delta(t) + \frac{[x-x_\mathrm{G}(t)]^2}{2a}\ .
\end{equation}
Moreover, the viscous flow can be described within the lubrication theory. The fields of the problem are the horizontal fluid velocity field $u(x,z,t)$, the vertical fluid velocity field $v(x,z,t)$, and the excess hydrodynamic pressure field $p(x,z,t)$ with respect to the atmospheric pressure, together with the local ionic concentrations $n_{\pm}(x,z,t)$, and the electric potential. The latter is decomposed into the sum of a quasi-static electric potential $\phi_0(x,z,t)$, and a streaming potential $\phi_1(x,t)$ that we assume to be $z$-invariant within the lubrication approximation~\cite{Zhao2020} (see Appendix for details). We further stress that the potential $\phi_0(x,z,t)$ is evaluated for the instantaneous geometry, and time enters the calculations parametrically.

Let us now non-dimensionalize the problem, through: 
\begin{equation}
\begin{aligned}
z &= a\epsilon Z\ , & x &= a\sqrt{2\epsilon}\,X\ , & \theta &= \sqrt{2\epsilon}\,\Theta\ ,\\
h &= a\epsilon H\ , & x_{\mathrm{G}} &= a\sqrt{2\epsilon}\,X_{\mathrm{G}}\ , &\delta &= a\epsilon \Delta\ , \\
t &= \frac{a\sqrt{2\epsilon}}{c}T\ , & u &= cU\ , & v &= c\sqrt{\frac{\epsilon}{2}}V\ ,\\
p &= \frac{\sqrt{2}\eta c}{a}\epsilon^{-3/2}P\ , & n_{\pm} &= n_{\infty}N_{\pm}\ , & \phi_0 &= \frac{k_{\textrm{B}} \mathcal{T}}{\tilde{z}e}\Phi_0\ ,\\
\phi_1 &= \frac{k_{\textrm{B}} \mathcal{T}}{\tilde{z}e}\Phi_1 , & \zeta_\textrm{p} &= \frac{k_{\textrm{B}} \mathcal{T}}{\tilde{z}e}\Xi_\textrm{p}\ , & \zeta_\textrm{w} &= \frac{k_{\textrm{B}} \mathcal{T}}{\tilde{z}e}\Xi_\textrm{w}\ ,
\end{aligned}
\label{adim}
\end{equation}
where $c$ is a typical horizontal fluid velocity scale, $e$ is the elementary electric charge, $k_{\textrm{B}}$ is the Boltzmann constant, and $\mathcal{T}$ is the constant temperature, assuming isothermal conditions throughout. We stress that we have arbitrarily used here the same scale for the two electric potentials $\phi_0$ and $\phi_1$. This will be naturally compensated by a supplementary dependency in $\textrm{Pe}/\sqrt{\epsilon}$, where the P\'eclet number is defined as $\textrm{Pe}=\sqrt{2}a\epsilon c/D$, within the dimensionless relation between $\Phi_0$ and $\Phi_1$, as we will see later. 

We have neglected any magnetic contribution because a scaling analysis based on Ohm's law and Ampere's law shows that the magnetic component of the Lorentz force is negligible compared with the electrostatic force under the present electrokinetic conditions. At leading order in $\epsilon$, the dimensionless steady incompressible Stokes equations describing the hydrodynamic flow under the volumic electric force read: 
\begin{align}
\frac{\partial^2 U}{\partial Z^2} &= \frac{\partial P}{\partial X} - \frac{\textrm{Ha}}{K^2} \frac{\partial^2 \Phi_0}{\partial Z^2} \, \left( \frac{\partial \Phi_0}{\partial X} + \frac{\partial \Phi_1}{\partial X} \right), \label{eqxmomentum}\\
\frac{\partial P}{\partial Z} &= \frac{\textrm{Ha}}{K^2} \frac{\partial^2 \Phi_0}{\partial Z^2} \, \frac{\partial \Phi_0}{\partial Z}\ , \label{eqzmomentum}
\end{align}
where $K = a\epsilon \kappa$ is the dimensionless version of the inverse Debye length $\kappa=\sqrt{2 \tilde{z}^2 e^2 n_{\infty}/(\varepsilon k_{\textrm{B}} \mathcal{T}})$, with $\varepsilon$ the permittivity of the medium, and $\textrm{Ha}= \sqrt{2} \epsilon^{3/2} a n_{\infty} k_{\textrm{B}} \mathcal{T}/(c \eta)$ is the Hartman number, which is a dimensionless parameter measuring the relative importance of electrostatic forces over viscous ones. We further stress that we also assumed  $\epsilon^2\partial_X^2\Phi_1\ll\partial_Z^2\Phi_0$, which will be justified later on since our set of assumptions implies $\epsilon^{3/2} \textrm{Pe}\ll1$.

Let us now turn to the description of ionic transport. We invoke here the dimensionless Nernst-Planck equation, describing the conservation of ionic matter under diffusion as well as advection by the combination of the flow and the electric force:
\begin{equation}
\label{nernst}
\begin{aligned}
\sqrt{\epsilon} \textrm{Pe} &\Bigg[\frac{\partial N_{\pm}}{\partial T} +\frac{\partial (N_{\pm}U)}{\partial X}+\frac{\partial (N_{\pm}V)}{\partial Z}\Bigg]\\ =&\epsilon \Bigg\{\frac{\partial^2 N_{\pm}}{\partial X^2} \pm \frac{\partial}{\partial X}\Bigg[N_{\pm}\Bigg( \frac{\partial \Phi_0}{\partial X}+\frac{\partial \Phi_1}{\partial X}\Bigg)\Bigg] \Bigg\}\\  &+ 2\frac{\partial^2 N_{\pm}}{\partial Z^2} \pm 2\frac{\partial}{\partial Z}\left(N_{\pm} \frac{\partial \Phi_0}{\partial Z}\right)\ .
\end{aligned}
\end{equation}

In the following, in addition to $\epsilon\ll1$, we assume that $\sqrt{\epsilon} \textrm{Pe}\ll1$. Thus, the maximum admissible $\textrm{Pe}$ scales as $\mathrm{Pe}\ll\epsilon^{-1/2}$, implying that the range of allowed advection magnitudes in our model increases as the lubrication gap becomes narrower. Writing Eq.~(\ref{nernst}) at leading order, and imposing no-flux boundary conditions at the solid-liquid interfaces, as well as $N_{\pm}\rightarrow 1$ and $\Phi_0\rightarrow 0$ at $X\rightarrow \pm \infty$, leads to the equilibrium Boltzmann factors $N_{\pm}=\exp(\mp\Phi_0)$. Injecting the latter into the leading-order dimensionless Poisson equation:
\begin{equation}
\frac{\partial^2\Phi_0}{\partial Z^2}=\frac{K^2}{2}(N_{-}-N_{+})\ ,
\end{equation}
and assuming $\Phi_0\ll1$, leads to the Debye-H\"uckel equation:
\begin{align}
\frac{\partial ^2 \Phi_0}{\partial Z^2}=K^2 \Phi_0\ . 
\label{chargeDist}
\end{align}

The latter can be solved using the two dimensionless surface electric potentials:
\begin{align}
\Phi_0|_{Z=0}=\Xi_{\textrm{w}} ~ \text{and}~ \Phi_0|_{Z=H}=\Xi_{\textrm{p}}\ . 
\label{chargeBc}
\end{align}

The Debye--H\"uckel approximation relies mainly on the linearization condition $|\Xi_{\textrm{P,W}}| = |\tilde z e \zeta_{\textrm{P,W}}|/(k_{\textrm{B}} T)\ll 1$, which corresponds to surface potentials of order $k_{\textrm{B}} T/(\tilde z e)\approx 25\,\mathrm{mV}$ or smaller for monovalent electrolytes. In addition, the present framework relies on a mean-field description and therefore neglects ion-ion correlations and other non-linear effects. A standard measure of the correlation strength is the electrostatic coupling parameter $\Gamma=\tilde z^2 e^2/(4\pi\varepsilon a k_{\textrm{B}} T)=\tilde z^2 l_{\textrm{B}}/a$, where $l_{\textrm{B}}$ is the Bjerrum length and $a\sim n_\infty^{-1/3}$ is the characteristic interionic spacing. Physically, $\Gamma$ represents the ratio of the Coulomb interaction energy and the thermal energy. The mean-field Debye-H\"uckel and Nernst-Planck descriptions are expected to remain reliable when $\Gamma\ll1$. Furthermore, the continuum description of ionic transport under confinement remains valid provided that $h_{\textrm{min}}\gg a$, where $h_{\textrm{min}}$ represents the minimum gap between the cylinder and wall. The number of ions contained within a characteristic gap volume scales as $N\sim n_\infty h_{\textrm{min}}^3=(h_{\textrm{min}}/a)^3$, so that $N\gg1$ when $h_{\textrm{min}}\gg a$. In this regime, many ions would populate the confined region and the ionic concentration can be treated as a continuum field.

Now, calculating the derivatives of Eq.~(\ref{eqxmomentum}) with respect to $Z$ and Eq.~(\ref{eqzmomentum}) with respect to $X$, adding the two obtained equations, and invoking Eq.~({\ref{chargeDist}), leads to:
\begin{align}
\frac{\partial^3 U}{\partial Z^3}~ +  \textrm{Ha} ~\frac{\partial \Phi_1}{\partial X}~\frac{\partial \Phi_0}{\partial Z}=0\ . 
\label{eqcombinedeq}
\end{align}
We further impose no-slip boundary conditions at the solid-liquid interfaces, and vanishing of the hydrodynamic pressure in the far field, through:
\begin{align}
\quad U|_{Z=0} = 0\ , U|_{Z=H}= \dot{X}_{\text{G}} + \dot{\Theta}\ , P|_{X\rightarrow \pm\infty} \rightarrow 0\ .
\label{BCs}
\end{align}
In addition, mass conservation can be expressed as:
\begin{align}
\frac{\partial H}{\partial T}+\frac{\partial}{\partial X}\int_0^H\mathrm{d}Z\, U =0\ . 
\label{volconv}
\end{align}
All together, Eqs.~(\ref{eqxmomentum}, \ref{chargeDist}, \ref{chargeBc}, \ref{eqcombinedeq}, \ref{BCs}, \ref{volconv}) allow to find a relation between the field $U$ and the field $\Phi_1$. This relation also involves the trajectory of the cylinder through $\Delta(T)$, $X_{\textrm{G}}(T)$, and $\Theta(T)$, as well as the dimensionless parameters Ha, $K$, $\Xi_{\textrm{p}}$ and $\Xi_{\textrm{w}}$. 

To close the problem, one needs another relation between $U$ and $\Phi_1$. Let us then consider the two longitudinal ionic fluxes $j_{x,\pm}=n_{\infty}c\ ,J_{X,\pm}$, where $J_{X,\pm}$ are the corresponding dimensionless ionic fluxes. Combining flow advection, Fick's law, and the migration due to the electric force, one has:
\begin{equation}
\begin{aligned}
J_{X,\pm} = U \ N_{\pm} - \frac{\sqrt{\epsilon}}{\textrm{Pe}} \left[ \frac{\partial N_{\pm}}{\partial X}\pm N_{\pm}\left(\frac{\partial \Phi_0}{\partial X}+\frac{\partial \Phi_1}{\partial X}\right)\right]\ ,
\end{aligned}
\end{equation}
and thus the current density:
\begin{align}
J_{X,+}-J_{X,-}= -2 U \Phi_0 - \frac{2 \sqrt{\epsilon}}{\textrm{Pe}} \frac{\partial \Phi_1}{\partial X}\ ,
\end{align}
where we invoked the previous Boltzmann factors for $N_{\pm}$, but retaining only the linear order in $\Phi_0$. Using the condition of zero net current in the fluid gap, \textit{i.e.} $\int_0^H\mathrm{d}Z\, (J_{X,+}-J_{X,-})=0$, one obtains:
\begin{align}
\frac{\partial \Phi_1}{\partial X} &= -\frac{\textrm{Pe}}{\sqrt{\epsilon}H} \int_0^H \mathrm{d}Z\,U \Phi_0\ .
\label{strm}
\end{align}
From the latter relation, we see that the gradient of the streaming potential is given by the gap-averaged value of the product between the static potential and the fluid velocity. We also get the Pe factor between $\Phi_1$ and $\Phi_0$, as announced after Eq.~(\ref{adim}).

Finally, combining Eqs.~(\ref{eqxmomentum}, \ref{chargeDist}, \ref{chargeBc}, \ref{eqcombinedeq}, \ref{BCs}, \ref{volconv}, \ref{strm}), one gets the explicit expression of the dimensionless gradient of the streaming potential:
\begin{widetext}
\begin{equation}
\frac{\partial \Phi_1}{\partial X} = \frac{-\frac{\textrm{Pe}}{\sqrt{\epsilon}H}\Big\{\Big(\frac{\dot{X}_{\text{G}}+\dot{\Theta}}{H}\Big)\mathcal{L}+\frac{6}{H^3} \Big[-\dot{\Delta}X +(X-X_{\text{G}})^2 \dot{X}_{\text{G}} -(\dot{X}_{\text{G}}+\dot{\Theta})\frac{H}{2}\Big]\mathcal{N}-\frac{6}{H^3}\mathcal{P}\mathcal{N}\Big\}}{1+\frac{\textrm{Pe}}{\sqrt{\epsilon}H}\Big\{ \frac{\textrm{Ha}}{K^2}\mathcal{M}-\frac{6}{H^3}\frac{\textrm{Ha}}{K^2}(\Xi_{\textrm{p}}+\Xi_{\textrm{w}})\Big[ \frac{H}{2}-\frac{1}{K}\tanh{\left(\frac{KH}{2}\right)}\Big]\mathcal{N}\Big\}}\ ,\label{streampot}
\end{equation}
\end{widetext}
where we introduced the following auxiliary functions for convenience:
$\mathcal{L}=\int_0^H\mathrm{d}Z\, \Phi_0Z,~\mathcal{M}=\int_0^H\mathrm{d}Z\, [\Xi_{\textrm{w}}-(\Xi_{\textrm{w}}-\Xi_{\textrm{p}})(Z/H)-\Phi_0]\Phi_0,~\mathcal{N}=\int_0^H\mathrm{d}Z\, \Phi_0Z(H-Z)$, and $\mathcal{P}$ as provided in the Appendix. Also, for convenience, the main variables, their definitions, and the dimensionless variables are summarised in Tab. \ref{tablenot}.
\begin{table}
\centering
\begin{tabular}{|lll|}
\hline
Symbol & Description & Definition / Scaling \\
\hline
$X,Z$ & Dimensionless coordinates & $x=a\sqrt{2\epsilon}\,X$, \;\; $z=a\epsilon Z$ \\
$T$ & Dimensionless time & $t=(a\sqrt{2\epsilon}/c)\,T$ \\
$H$ & Dimensionless gap profile & $h=a\epsilon H$ \\
$\Delta$ & Dimensionless minimum gap & $\delta=a\epsilon \Delta$ \\
$X_{\mathrm{G}}$ & Dimensionless cylinder position & $x_{\mathrm{G}}=a\sqrt{2\epsilon}\,X_{\mathrm{G}}$ \\
$\Theta$ & Dimensionless rotation angle & $\theta=\sqrt{2\epsilon}\,\Theta$ \\
$U,V$ & Dimensionless fluid velocities & $u=cU$, \;\; $v=c\sqrt{\epsilon/2}\,V$ \\
$P$ & Dimensionless pressure & $p=(\sqrt{2}\eta c/a)\epsilon^{-3/2}P$ \\
$N_\pm$ & Dimensionless ion concentrations & $n_\pm=n_\infty N_\pm$ \\
$\Phi_0$ & Quasi-static electric potential & $\phi_0=(k_{\textrm{B}}T/\tilde z e)\phi_0$ \\
$\Phi_1$ & Streaming potential & $\phi_1=(k_{\textrm{B}}T/\tilde z e)\phi_1$ \\
$\Xi_{\textrm{p}}$ & Cylinder surface potential & $\zeta_{\textrm{p}}=(k_{\textrm{B}}T/\tilde z e)\Xi_{\textrm{p}}$ \\
$\Xi_{\textrm{w}}$ & Wall surface potential & $\zeta_{\textrm{w}}=(k_{\textrm{B}}T/\tilde z e)\Xi_{\textrm{w}}$ \\
Pe & P\'eclet number & Pe$=\sqrt{2}\,a\epsilon c/D$ \\
$K$ & Dimensionless inverse Debye length & $K=a\epsilon\kappa$ \\
Ha & Hartmann number & Ha$=\sqrt{2}\,\epsilon^{3/2} a n_\infty k_{\textrm{B}} T/(c\eta)$ \\
$\xi$ & Dimensionless particle inertia & $6\eta/(\epsilon a \rho c)$ \\
\hline
\end{tabular}
\caption{The principal variables, their definitions and their dimensionless versions.}
\label{tablenot}
\end{table}

\section{Results and Discussions}
From the integrals over $X$ of the dominant lubrication pressure, viscous stress, and Maxwell's stress, one can then obtain the dominant dimensionless forces and torques per unit length exerted on the cylinder~\cite{SalezMahadevan2015}, as detailed in the Appendix. Then, the conservation of linear and angular momenta leads to the following equations of motion:
\begin{widetext}
\begin{align}
\frac{\mathrm{d}^2 \Delta}{\mathrm{d}T^2}
&= F_{\perp}+\frac{2 \xi}{3 \pi}
\int_{-\infty}^{\infty}
\mathrm{d}X\left[\,
P- \frac{\mathrm{Ha}}{2K^2}
\Big(\frac{\partial \Phi_0}{\partial Z}\Big)^2
+ \frac{\mathrm{Ha}}{4K^2}\epsilon
\Big(\frac{\partial \Phi_1}{\partial X}\Big)^2\right]_{Z=H}
\ , 
\label{eqperp}\\[6pt]
\frac{\mathrm{d}^2 X_{\text{G}}}{\mathrm{d}T^2}
&= F_{\parallel}-\frac{2 \epsilon \xi}{3 \pi}
\int_{-\infty}^{\infty}
\mathrm{d}X\,\Bigg[
(X-X_{\text{G}}) P
+ \frac{1}{2}\frac{\partial U}{\partial Z}
-\frac{\mathrm{Ha}}{4K^2}\epsilon (X-X_{\text{G}})
\Big(\frac{\partial \Phi_1}{\partial X}\Big)^2 + \frac{\mathrm{Ha}}{2K^2} \frac{\partial \Phi_1}{\partial X} \frac{\partial \Phi_0}{\partial Z}
\Bigg]_{Z=H}\ , \label{eqpara}\\[6pt]
\frac{\mathrm{d}^2 \Theta}{\mathrm{d}T^2}
&= \frac{4 \epsilon \xi}{3 \pi}
   \int_{-\infty}^{\infty}\mathrm{d}X\,\Bigg[
   -\frac{1}{2}\frac{\partial U}{\partial Z}\,
   + \frac{\mathrm{Ha}}{2K^2}\epsilon (X-X_{\textrm{G}})
     \Big(\frac{\partial \Phi_1}{\partial X}\Big)^2  -\frac{\mathrm{Ha}}{2K^2} \frac{\partial \Phi_1}{\partial X} \frac{\partial \Phi_0}{\partial Z}     
     \Bigg]_{Z=H}\ , \label{eqrot}
\end{align}
\end{widetext}
where we introduced the dimensionless viscosity $\xi =6 \eta/(\epsilon a \rho c)$, as well as dimensionless external normal and longitudinal forces per unit length, $F_{\perp}$ and $F_{\parallel}$, or their dimensional versions $f_{\perp}=\pi a \rho c^2F_{\perp}/2$ and $f_{\parallel}=\pi a\rho c^2F_{\parallel}/\sqrt{2\epsilon}$, respectively. To illustrate the dynamics of these three coupled non-linear equations of motion, we successively address below three specific motions with an increasing number of degrees of freedom.

\begin{figure}[t!]
    \centering
    \includegraphics[width=0.6\linewidth]{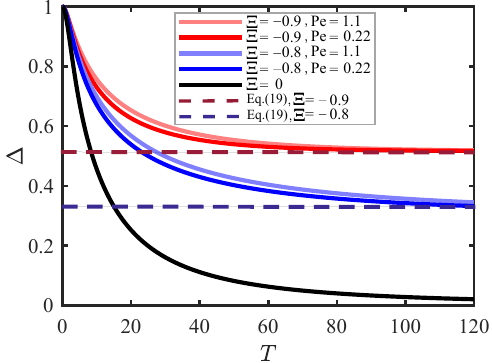}
\caption{\textit{Dimensionless distance $\Delta$ between the cylinder and the wall a as function of the dimensionless time $T$, as obtained from numerically solving Eq.~(\ref{eqperp}), with $\Delta(0) = 1$ and $\dot{\Delta}(0) = 0$, and with $\Theta(T)$ and $X_{\text{G}}(T)$ identically set to $0$. The common fixed parameters are $K=2$, $\mathrm{Ha}=1$, $F_{\perp} = -0.1$, $F_{\parallel} = 0$, $\epsilon =0.05$, and $\xi$ =1, while Pe and $ \Xi\, (=\Xi_{\mathrm{p}} = \Xi_{\mathrm{w}})$ are varied as indicated by the colour code. The dashed lines correspond to the solutions of Eq.~(\ref{sedim}). }}
    \label{fig2}
\end{figure}
\begin{figure}[t!]
    \centering
    \includegraphics[width=0.32\columnwidth]{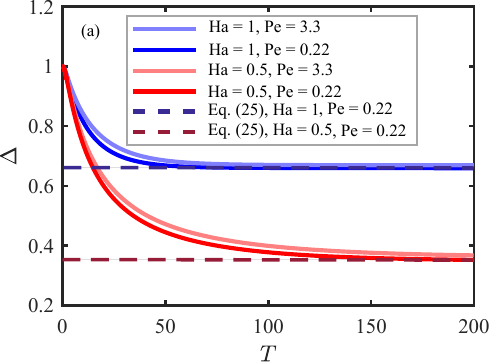}   
    \includegraphics[width=0.31\columnwidth]{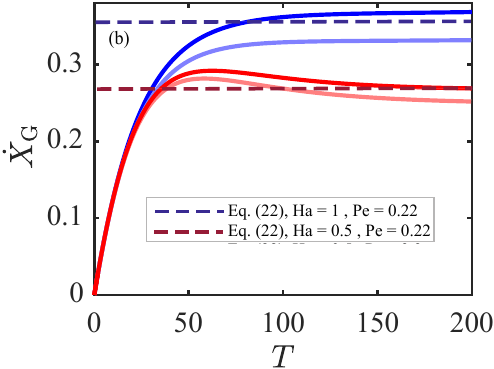}
    \includegraphics[width=0.32\columnwidth]{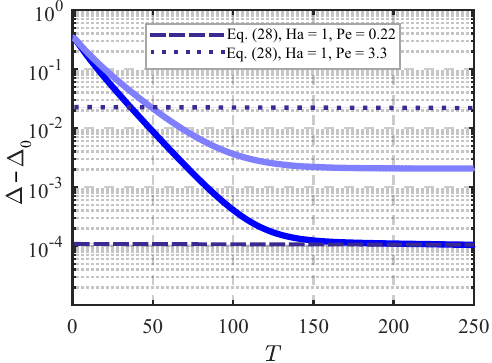}
\caption{\textit{(a) Dimensionless distance $\Delta$ between the cylinder and the wall as a function of the dimensionless time $T$. The dashed lines correspond to the solutions of Eq.~(\ref{eqperpcoupled2}). (b) Dimensionless longitudinal velocity $\dot{X}_{\textrm{G}}$ of the cylinder as a function dimensionless time $T$. The dashed lines correspond to the solutions of Eq.~(\ref{slidingvel}). (c) Difference $\Delta - \Delta_0$, between the dimensionless distance $\Delta$ and the pure-sedimentation equilibrium one $\Delta_0$ obtained from Eq.~(\ref{sedim}), as a function of dimensionless time $T$. The curves in all panels are obtained from numerically solving Eqs.~(\ref{eqperp}-\ref{eqpara}), with $\Delta(0) = 1$, $X_{\textrm{G}}(0)=0$, $\dot{\Delta}(0) = \dot{X}_{\textrm{G}}(0)=0$, and with $\Theta(T)$ identically set to 0. The common fixed parameters are $K=3$, $F_{\perp} = -0.05$, $F_{\parallel} = 0.015$, $\epsilon =0.05$, $\Xi_{\mathrm{p}} = \Xi_{\mathrm{w}}=-1$, and $\xi =1$, while Pe and Ha are varied as indicated by the colour code. }}
    \label{Constantforce}
\end{figure}
\begin{figure}
    \centering
    \includegraphics[width=0.32\columnwidth]{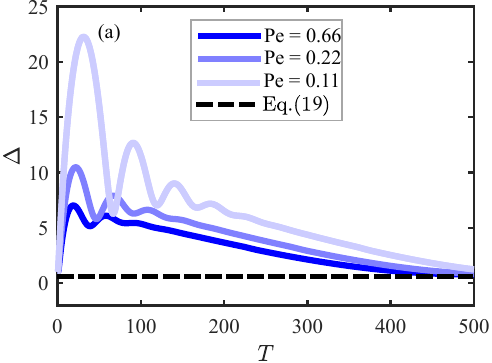}  
    \includegraphics[width=0.32\columnwidth]{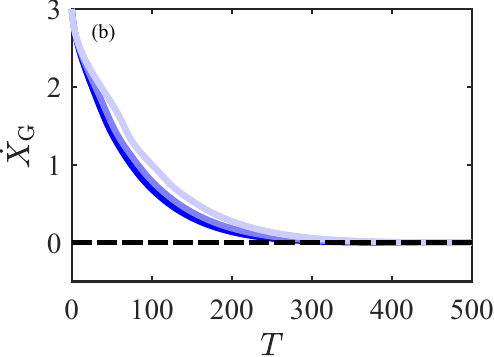}
    \includegraphics[width=0.32\columnwidth]{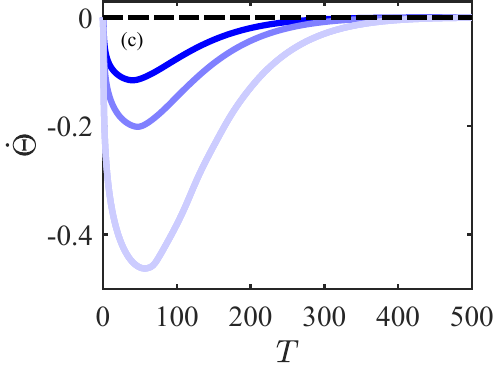}
\caption{\textit{(a) Dimensionless distance $\Delta$ between the cylinder and the wall as a function of the dimensionless time $T$. The dashed line corresponds to the solution of Eq.~(\ref{sedim}). (b) Dimensionless longitudinal velocity $\dot{X}_{\textrm{G}}$ of the cylinder as a function dimensionless time $T$. (c) Dimensionless angular velocity $\dot{\Theta}$ of the cylinder as a function dimensionless time $T$. The curves in all panels are obtained from numerically solving Eqs.~(\ref{eqperp}-\ref{eqrot}), with $\Delta(0) = 1$, $X_{\textrm{G}}(0)=0$, $\Theta(0)=0$, $\dot{X}_{\textrm{G}}(0)=3$, and $\dot{\Delta}(0) =\dot{\Theta}(0)= 0$. The common fixed parameters are $\textrm{Ha}=1$, $K=3$, $F_{\perp} = -0.04$, $F_{\parallel} = 0$, $\epsilon =0.05$, $\Xi_{\mathrm{p}} = \Xi_{\mathrm{w}}=-1$, and $\xi =1$, while Pe is varied as indicated by the colour code.}}
    \label{free}
\end{figure}

\subsection{Sedimentation}
We first consider a pure normal motion, where the variables $\Theta$ and $X_{\text{G}}$ are identically set to $0$. Numerical solutions are presented in Fig.~\ref{fig2}. In the case of uncharged cylinder and wall, the gap $\Delta$ decreases gradually towards zero under the influence of the constant dimensionless gravitational-like force per unit length $F_{\perp}$. This motion is slowed down by the drag force generated by the lubrication pressure. However, when we consider charged cylinder and wall, the electrokinetic and electrostatic repulsion forces also counteract the action of $F_{\perp}$. In the transient regime, the larger $\mathrm{Pe}$ is, the more electrokinetic repulsion there is. At long times, when the cylinder's velocity vanishes, so does the electrokinetic repulsive force. Hence, the equilibrium distance between the cylinder and the wall is solely determined by the balance between the gravitational and electrostatic forces, as:
\begin{equation}
\begin{aligned}
\frac{\xi\mathrm{Ha}}{3 \pi}  \int_{-\infty}^{\infty} \mathrm{d}X\,\mathcal{I}(X,\Delta) + F_{\perp}=0\ ,
\label{sedim}
\end{aligned}
\end{equation}
where:
\begin{equation}
\begin{aligned}
\mathcal{I}(X, \Delta) =  \frac{-\Xi_{\textrm{p}}^2 -\Xi_{\textrm{w}}^2 + 2 \ \Xi_{\textrm{p}} \ \Xi_{\textrm{w}} \cosh{[K(\Delta + X^2)]}}{\sinh^2{[K(\Delta+X^2)]}} \ .
\label{aux1}
\end{aligned}
\end{equation}
As shown in Fig.~\ref{fig2}, Eq.~(\ref{sedim}) captures well the numerical solutions at long times.

\subsection{Sliding}
We turn to the coupling between normal and longitudinal degrees of freedom. To do so, the cylinder is now also pulled with a constant dimensionless longitudinal force per unit length $F_{\parallel}$. Numerical solutions are presented in Fig.~\ref{Constantforce}. Qualitatively, the normal motion in Fig.~\ref{Constantforce}(a) ressembles the pure sedimentation one in Fig.~\ref{fig2}, with a Pe-dependent transient regime and a longterm equilibrium-like regime. However, there are crucial differences that we investigate in the following.

At long times, the cylinder reaches a non-zero terminal longitudinal velocity. The distance between the cylinder and the wall having reached its steady value, Eq. (\ref{streampot}) loses its fore-aft asymmetry about $X=X_{\mathrm{G}}$, causing the corresponding electrokinetic contribution to the longitudinal force to vanish. Therefore, the electrokinetic term $\frac{\mathrm{Ha}}{4K^2}\epsilon (X-X_{\mathrm{G}})\left(\frac{\partial \Phi_1}{\partial X}\right)^2$ becomes an odd function of $X-X_{\mathrm{G}}$ and yields $0$ upon integration. The steady longitudinal regime is thus given by the force balance:
\begin{equation}
\begin{aligned}
   \frac{2 \epsilon \xi}{3 \pi}
   \int_{-\infty}^{\infty}\mathrm{d}X\,
   \left[
   (X-X_{\text{G}}) P
   + \frac{1}{2}\frac{\partial U}{\partial Z}+\frac{\mathrm{Ha}}{2K^2} \frac{\partial \Phi_1}{\partial X} \frac{\partial \Phi_0}{\partial Z}\right]_{Z=H}  = F_{\parallel} \ .\label{eqparacoupled2}  
\end{aligned}
\end{equation}
In the particular case where $\frac{\mathrm{Ha}}{K^2} \frac{\mathrm{Pe}}{\sqrt{\epsilon}} \ll 1$ and $K \gg 1$, we can further simplify Eq.~(\ref{eqparacoupled2}), as:
\begin{equation}
\begin{aligned}
 \frac{\epsilon \xi}{3 \pi} \Big\{
\frac{2\pi}{\Delta^{1/2}} 
+ \frac{\mathrm{Ha}}{K^2}\frac{\mathrm{Pe}^2}{\epsilon} \Big[ 
\frac{3\pi \Xi_{\textrm{p}}(\Xi_{\textrm{p}} + \Xi_{\textrm{w}})}{4K\Delta^{3/2}}
+\frac{3\pi (\Xi_{\textrm{w}}^2 - \Xi_{\textrm{p}}^2)}{8K^2\Delta^{5/2}}
+ \frac{3\pi(\Xi_{\textrm{p}} + \Xi_{\textrm{w}})^2}{16K^2\Delta^{5/2}}\Big]\\ -  
\frac{\mathrm{Ha}}{K^2}\frac{\mathrm{Pe}}{\sqrt{\epsilon}}\Big(-2 \Delta f_1 + \frac{2}{3}\Delta^2 f_4\Big) - \frac{\mathrm{Ha}}{K^2}\frac{\mathrm{Pe}}{\sqrt{\epsilon}}\Big[\frac{\pi \Xi_{\textrm{p}}(\Xi_{\textrm{p}}-\Xi_{\textrm{w}})}{2K\Delta^{3/2}} + \frac{3\pi\Xi_{\textrm{p}}(\Xi_{\textrm{p}}+\Xi_{\textrm{w}})}{2K\Delta^{3/2}}-\frac{\pi \Xi_{\mathrm{p}}^2}{\Delta^{1/2}}\Big]\\+
\frac{\mathrm{Ha}}{K^2}\frac{\mathrm{Pe}}{\sqrt{\epsilon}} \Big[\frac{\pi \Xi_{\mathrm{p}}^2}{\Delta^{1/2}} + \pi \frac{\Xi_{\mathrm{p}}(\Xi_{\mathrm{w}}-\Xi_{\mathrm{p}})}{\Delta^{3/2}} \Big] 
\Big\} \dot{X_{\text{G}}} \simeq F_{\parallel}\ , \label{slidingvel}
\end{aligned}
\end{equation}
where the complete forms of $f_1$ and $f_4$ are provided in the Appendix, and their asymptotic values at long times are given by:
\begin{equation}
\begin{aligned}
    f_1 \simeq \frac{\mathrm{Pe}}{\sqrt{\epsilon}} (\Xi_{\textrm{p}} + \Xi_{\textrm{w}}) \Big[ -\frac{3\pi \Xi_{\textrm{p}}}{8K\Delta^{5/2}} - \frac{\Xi_{\textrm{w}} -\Xi_{\textrm{p}}}{K^2}\frac{5\pi}{16 \Delta^{7/2}} +  \frac{6(\Xi_{\textrm{p}}+\Xi_{\textrm{w}})}{K^2}\frac{15 \pi}{128 \Delta^{7/2}}\Big] \ ,
\end{aligned}
\end{equation}
\begin{equation}
\begin{aligned}    
f_4 &\simeq \frac{\mathrm{Pe}}{\sqrt{\epsilon}} \frac{(\Xi_{\textrm{p}} + \Xi_{\textrm{w}})^2}{K^2} \frac{35 \pi}{128 \Delta^{9/2}}\ .
\end{aligned}
\end{equation}
As shown in Fig.~\ref{Constantforce}(b), Eq.~(\ref{slidingvel}) captures well the steady regime reached by the numerical solution for the smallest Ha value, while, at higher $\mathrm{Ha}$, we see a departure of the numerical solution with respect to the prediction of Eq.~(\ref{slidingvel}).

Let us now have a closer look at the normal part of the motion, in Fig.~\ref{Constantforce}(a). The long-term steady distance between the cylinder and the wall increases with Ha. This is partly resulting from the electrostatic force. However, unlike the pure-sedimentation case of Fig.~\ref{fig2}, the longitudinal motion of the cylinder induces an additional normal electroviscous lift force. Hence, and since one can show that the pressure contribution vanishes in the steady state, the steady normal distance is now given by the solution of:
\begin{equation}
\begin{aligned}
    \frac{\xi\mathrm{Ha}}{3 \pi}\int_{-\infty}^{\infty} \mathrm{d}X\,\mathcal{I}(X, \Delta) +    F_{\textrm{ev}} 
 + F_{\perp}=0\ . \label{eqperpcoupled2}
\end{aligned}
\end{equation}
where we introduced the dimensionless electroviscous lift force per unit length:
\begin{equation}
   F_{\textrm{ev}} = \frac{\xi\mathrm{Ha}\epsilon}{6 \pi K^2} \int_{-\infty}^{\infty}   \mathrm{d}X \,
    \Bigg[ \Big(\frac{\partial \Phi_1}{\partial X}\Big)^2\Bigg]_{T\rightarrow\infty}\ . \label{Elec1}
\end{equation}
As shown in Fig.~\ref{Constantforce}(a), Eq.~(\ref{eqperpcoupled2}) captures well the steady regime reached by the numerical solution. 

Furthermore, in the particular case where $\frac{\mathrm{Ha}}{K^2} \frac{\mathrm{Pe}}{\sqrt{\epsilon}}\ll 1$ and $K \gg 1$, we can simplify Eq.~(\ref{Elec1}), as:
\begin{equation}
F_{\textrm{ev}} \simeq \frac{\xi}{6}\frac{\mathrm{Ha}}{K^2}\mathrm{Pe^2}\dot{X}^2_{\mathrm{G},\infty}\Bigg[\frac{\Xi_{\mathrm{p}}^2}{2K^2\Delta_\infty^{3/2}}+\frac{3\Xi_{\mathrm{p}}(\Xi_{\mathrm{w}}-\Xi_{\mathrm{p}})}{4K^3\Delta_\infty^{5/2}}-\frac{\Xi_{\mathrm{p}}(\Xi_{\mathrm{w}}+\Xi_{\mathrm{p}})}{4K^3\Delta_\infty^{5/2}} \Bigg]\ ,
\label{lift1}
\end{equation}
where $\dot{X}_{\mathrm{G},\infty}$ and $\Delta_\infty$ denote the steady values of $\dot{X}_{\mathrm{G}}$ and $\Delta$, respectively.
Self-consistently expanding the steady normal distance $\Delta_\infty$, around the equilibrium gap $\Delta_0$ reached due to the competition between pure electrostatics and gravity (see Eq.~(\ref{sedim})),  \textit{i.e.} $\Delta_{\infty} \simeq \Delta_{0} + \frac{\mathrm{Ha}}{K^2} \frac{\mathrm{Pe}}{\sqrt{\epsilon}} \Delta_{1}$, and invoking Eqs. (\ref{eqperpcoupled2}) and~(\ref{lift1}), one gets :
\begin{equation}
    \Bigg\{\frac{\xi \mathrm{Ha} K \Xi^2}{3 \pi}  \int_{-\infty}^{\infty} \textrm{d}X \ \frac{\tanh{[\frac{K}{2}(\Delta_{0}+X^2)]}}{\cosh^2{[\frac{K}{2}(\Delta_{0}+X^2)]}} +  \frac{\xi \mathrm{Ha} \mathrm{Pe}^2 \dot{X}^2_{\mathrm{G},\infty}}{8 K^4 \Delta_{0}^{5/2}}\Bigg\} \ \frac{\mathrm{Ha}\mathrm{Pe}}{K^2 \sqrt{\epsilon}} \Delta_{1} \simeq \frac{\xi \mathrm{Ha} \mathrm{Pe}^2 \dot{X}^2_{\mathrm{G},\infty}}{12 K^4 \Delta_{0}^{3/2}}\ ,
\end{equation}
in the particular case where $\Xi_\mathrm{p} = \Xi = \Xi_{\mathrm{w}}$. As shown in Fig.~\ref{Constantforce}(c), the previous asymptotic expression of the electroviscous lift force captures well the numerical solution at small Pe. Moreover, we recover the known fact that the magnitude of the asymptotic electroviscous lift is small~\cite{Bureau2023}. Besides, we find that the asymptotic expression overestimates the numerical lift force at higher $\mathrm{Pe}$.

Putting back dimensions in Eq.~(\ref{lift1}), through $f_{\textrm{ev}}=\pi a \rho c^2F_{\textrm{ev}}/2$, one gets the following asymptotic expression of the electroviscous lift force per unit length for a cylinder:
\begin{equation}
    \begin{aligned}
f_{\textrm{ev}} = \frac{\pi}{4\sqrt{2}} \frac{ \varepsilon^2k_{\textrm{B}} \mathcal{T} a^{1/2}\dot{x}_\mathrm{G,\infty}^2}{D^2n_\infty \tilde{z}^2e^2 \delta_\infty^{3/2}} \Bigg[ \zeta_{\mathrm{p}}^2 + \frac{3\zeta_{\mathrm{p}}(\zeta_{\mathrm{w}}-\zeta_{\mathrm{p}})}{2\kappa \delta_\infty}-\frac{\zeta_{\mathrm{p}}(\zeta_{\mathrm{w}}+\zeta_{\mathrm{p}})}{2\kappa\delta_\infty}\Bigg]\ ,
    \end{aligned}
    \label{dimres}
\end{equation} 
where $\dot{x}_{\mathrm{G},\infty}=c\dot{X}_{\mathrm{G},\infty}$ and $\delta_\infty=a\epsilon\Delta_\infty$. 
We stress that our asymptotic expression is different from the one obtained by Bike \textit{et al.}~\cite{Bike1990}, which is due to the fact that we have treated our system as an open circuit (no net current) while Bike \textit{et al.} have used the charge continuity condition. Our asymptotic expression is also different from the one obtained by Tabatabaei \textit{et al.}~\cite{Tabatabaei2006}, which is due to the fact that these authors have considered a diffusio-osmotic contribution, which happens to be a higher-order correction in our leading-order model. Finally, we note that the scaling of the electroviscous lift force for a sphere can be obtained by multiplying Eq.~(\ref{dimres}) by the hydrodynamic radius $\sim\sqrt{a \delta_{\infty}}$.

\subsection{Free motion}
In the last considered case, we combine the three degrees of freedom of the cylinder, through the coupled normal, longitudinal and rotational motions. To do so, we consider the situation where the cylinder experiences only the gravitational force per unit length $F_{\perp}$ as an external driving, but is released with a non-zero initial longitudinal velocity. Numerical solutions are shown in Fig.~\ref{free}. Figure \ref{free}(a) represents the vertical motion of the cylinder. At short times, the competition between gravity, electrokinetic forces and electrostatic forces leads to a damped oscillatory motion. The amplitude and period of the oscillations appear to increase with $\mathrm{Pe}$. Finally, at long times, the system reaches equilibrium, with no motion, as in the pure sedimentation case. 
As shown in Fig.~4(b), the longitudinal velocity exhibits a slow decay to zero, as the cylinder eventually reaches its equilibrium distance from the wall governed solely by gravity and electrostatics. Moreover, larger $\mathrm{Pe}$ values lead to a faster transient relaxation, indicating an intricate coupling mechanism with vertical motion and electrokinetics. 
A similar behaviour is observed for the angular velocity in Fig.~4(c). Although the cylinder is not subjected to any external torque, the coupling between translation, rotation, and electrohydrodynamic stresses generates a transient rotational motion. The magnitude of the induced rotation rate increases with $\mathrm{Pe}$, resulting in a breaking of the fore-aft symmetry of the lubrication flow. As the translational and normal motions gradually cease, the electrohydrodynamic torque also vanishes, and the cylinder eventually relaxes to its steady state, with $\dot\Theta\to0$.

\section{Conclusion}
We have developed a complete electrohydrodynamic lubrication theory for the free motion of a charged cylinder near a charged flat wall, within an ionic solution. Specifically, combining hydrodynamic lubrication theory, Debye-H\"uckel electrostatics,
and Nernst-Planck electrokinetics, we have derived the three coupled equations of motion for the normal, longitudinal and rotational degrees of freedom of the cylinder. As such, our results extend the classical Fax\'en-Brenner description of wall-corrected mobility in Stokes flow by incorporating surface charges, dissolved ions, and electroviscous effects. We then investigated the solutions of these equations by a combination of numerical resolutions and asymptotic analysis, through three canonical situations with an increasing number of degrees of freedom. Our results show the emergence of an electroviscous lift force, which becomes prominent at large P\'eclet numbers. While an asymptotic scaling is retrieved for such a lift force, as in previous works from the literature, our work allows to generalize it beyond usual approximations and with the intricate coupling of multiple degrees of freedom. Extending the current framework to the spherical case, adding complex boundaries, and accounting for thermal fluctuations in the colloidal motion~\cite{ye2025brownian} in some asymptotic regimes, would be relevant future developments.

\begin{acknowledgments}
T.S. met at several occasions Rudi Podgornik who was an invited professor at the Gulliver lab, ESPCI Paris, France. The broad scientific expertise and general cultural knowledge of Rudi were simply outstanding. By contributing to this topical issue on ``Statistical Physics: From Quantum Fluctuations to Soft Matter and Viral Assemblies", T.S. and his team express their admiration of Rudi Podgornik's work and celebrate his scientific heritage.

The authors thank Aditya Jha and Quentin Ferreira, for interesting discussions. They acknowledge financial support from the European Union through the European Research Council under EMetBrown (ERC-CoG-101039103) grant. They also acknowledge financial support from the Agence Nationale de la Recherche under EMetBrown (ANR-21-ERCC-0010-01), Softer (ANR21-CE06-0029), and Fricolas (ANR-21-CE06-0039) grants, as well as from the Interdisciplinary and Exploratory Research program under MISTIC grant at the University of Bordeaux, France. Besides, they acknowledge the support from the R\'eseaux de Recherche Impulsion Frontiers of Life which received financial support from the French government in the framework of the University of Bordeaux's France 2030 program. Finally, they thank the Soft Matter Collaborative Research Unit, Frontier Research Center for Advanced Material and Life Science, Faculty of Advanced Life Science at Hokkaido University, Sapporo, Japan, and the CNRS International Research Network between France and India on Hydrodynamics at small scales: From soft matter to bioengineering.
\end{acknowledgments}

\appendix
\section{Fluid velocity field}
\setcounter{equation}{0}
Equation~(\ref{eqcombinedeq}) has a solution of the form:
\begin{equation}
    \begin{aligned}
        U(X,Z,T)=-\frac{\mathrm{Ha}}{K^2} \frac{\partial \Phi_1}{\partial X} \Phi_0 -A(X,T)\frac{Z^2}{2}-B(X,T)Z
        -C(X,T)\ , \label{Uform}
    \end{aligned}
\end{equation}
where $A$, $B$ and $C$ are integration constants. Using the two velocity boundary conditions in Eq.~(\ref{BCs}), one further gets:
\begin{equation}
    \begin{aligned}
        U = (\dot{X}_{\textrm{G}} + \dot{\Theta})\frac{Z}{H} + \frac{\mathrm{Ha}}{K^2}  \frac{\partial \Phi_1}{\partial X}\Big[(\Xi_{\mathrm{p}} - \Xi_{\mathrm{w}})\frac{Z}{H} -\Phi_0 + \Xi_{\mathrm{w}}\Big]+ A(X,T) \frac{Z(H-Z)}{2}\ . \label{U1}
    \end{aligned}
\end{equation}
Then, integrating Eq.~(\ref{volconv}) with respect to $X$, we obtain:
\begin{equation}
   \mathcal{P}(T) + \int  \mathrm{d}X\,\frac{\partial H}{\partial T} + \int_0^{H}\mathrm{d}Z\,U  = 0\ , \label{volconv1}
\end{equation}
where $\mathcal{P}$ is an integration constant. Utilising Eqs.~(\ref{U1}, \ref{volconv1}), we obtain a relation between $A$ and $\mathcal{P}$, as: 
\begin{widetext}
    \begin{align}
        A(X) &= -\frac{12}{H^3}\Big\{ \dot{\Delta}X - (X-X_{\mathrm{G}})^2 \dot{X}_{\mathrm{G}} + (\dot{X}_{\mathrm{G}} + \dot{\Theta}) \frac{H}{2} +\frac{\mathrm{Ha}}{K^2}  \frac{\partial \Phi_1}{\partial X}(\Xi_{\mathrm{p}} + \Xi_{\mathrm{w}}) \Big[ \frac{H}{2} - \frac{1}{K}\tanh{\Big( \frac{KH}{2}\Big)}\Big] + \mathcal{P}\Big\}\ .
    \end{align}
\end{widetext}
From Eqs.~(\ref{eqxmomentum}, \ref{U1}), we can then write the longitudinal pressure gradient as:
\begin{align}
    \frac{\partial P}{\partial X} = \mathrm{Ha} \Phi_0 \frac{\partial \Phi_0}{\partial X}-A(X)\ . \label{gradp}
\end{align}
Finally, applying the vanishing pressure condition of Eq.~(\ref{BCs}), and invoking Eq.~(\ref{gradp}), we get:
\begin{widetext}
\begin{align}
 0 =\frac{9 \pi X_{\textrm{G}} \dot{\Delta}}{2 \Delta^{5/2}}-\frac{3 \pi \dot{X}_{\textrm{G}}}{2 \Delta^{3/2}}+\frac{3 \pi (\dot{X}_{\textrm{G}}+\dot{\Theta})}{\Delta^{3/2}}
    +\frac{9\pi \mathcal{P}}{2 \Delta^{5/2}}+6\frac{\mathrm{Ha}}{K^2} \underbrace{\int_{-\infty}^{\infty}\textrm{d}X\, \frac{1}{H^3} \frac{\partial \Phi_1}{\partial X} (\Xi_{\mathrm{p}} + \Xi_{\mathrm{w}}) \Big[ H -\frac{2}{K} \tanh{\Big( \frac{KH}{2}\Big)}\Big]}_{\dot{X}_{\textrm{G}} f_1+\dot{\Theta} f_2 + \dot{\Delta} f_3 + \mathcal{P} f_4}\ . \label{eqDT}
\end{align}
Lastly, to determine $\mathcal{P}$ explicitly, we rewrite the integral on the right-hand side of Eq.~(\ref{eqDT}) as:
\begin{align}
    \begin{aligned}
         \int_{-\infty}^{\infty}\textrm{d}X\, \frac{1}{H^3} \frac{\partial \Phi_1}{\partial X} (\Xi_{\mathrm{p}} + \Xi_{\mathrm{w}}) \Big[ H -\frac{2}{K} \tanh{\Big( \frac{KH}{2}\Big)}\Big] = \dot{X}_{\textrm{G}} f_1+\dot{\Theta} f_2 + \dot{\Delta} f_3 + \mathcal{P} f_4\ ,
    \end{aligned}
\end{align}
where:
\begin{equation}
    \begin{aligned}
        f_1 = \int_{-\infty}^{\infty}\textrm{d}X\, \frac{\frac{\mathrm{Pe}}{\sqrt{\epsilon}H^4} \Big\{ -\frac{\mathcal{L}}{H}-\frac{6}{H^3}\Big[(X-X_{\mathrm{G}})^2 - \frac{H}{2}\Big]\mathcal{N}\Big\}(\Xi_{\mathrm{p}}+\Xi_{\mathrm{w}})\Big[ H -\frac{2}{K} \tanh{\Big( \frac{KH}{2}\Big)}\Big]}{1+\frac{\textrm{Pe}}{\sqrt{\epsilon}H}\Big\{ \frac{\textrm{Ha}}{K^2}\mathcal{M}-\frac{6}{H^3}\frac{\textrm{Ha}}{K^2}(\Xi_{\textrm{p}}+\Xi_{\textrm{w}})\Big[ \frac{H}{2}-\frac{1}{K}\tanh{\left(\frac{KH}{2}\right)}\Big]\mathcal{N}\Big\}}\ ,
    \end{aligned}
\end{equation}
\begin{equation}
    \begin{aligned}
         f_2 = \int_{-\infty}^{\infty}\textrm{d}X\, \frac{\frac{\mathrm{Pe}}{\sqrt{\epsilon}H^4} \Big\{ -\frac{\mathcal{L}}{H}+\frac{3}{H^2}\mathcal{N}\Big\}(\Xi_{\mathrm{p}}+\Xi_{\mathrm{w}})\Big[H -\frac{2}{K} \tanh{\Big( \frac{KH}{2}\Big)}\Big]}{1+\frac{\textrm{Pe}}{\sqrt{\epsilon}H}\Big\{ \frac{\textrm{Ha}}{K^2}\mathcal{M}-\frac{6}{H^3}\frac{\textrm{Ha}}{K^2}(\Xi_{\textrm{p}}+\Xi_{\textrm{w}})\Big[ \frac{H}{2}-\frac{1}{K}\tanh{\left(\frac{KH}{2}\right)}\Big]\mathcal{N}\Big\}}\ ,
    \end{aligned}
\end{equation}
\begin{equation}
    \begin{aligned}
         f_3 = \int_{-\infty}^{\infty} \textrm{d}X\, \frac{\frac{\mathrm{Pe}}{\sqrt{\epsilon}H^4} \Big\{ \frac{6}{H^3} X \mathcal{N}\Big\}(\Xi_{\mathrm{p}}+\Xi_{\mathrm{w}})\Big[H -\frac{2}{K} \tanh{\Big( \frac{KH}{2}\Big)}\Big]}{1+\frac{\textrm{Pe}}{\sqrt{\epsilon}H}\Big\{ \frac{\textrm{Ha}}{K^2}\mathcal{M}-\frac{6}{H^3}\frac{\textrm{Ha}}{K^2}(\Xi_{\textrm{p}}+\Xi_{\textrm{w}})\Big[ \frac{H}{2}-\frac{1}{K}\tanh{\left(\frac{KH}{2}\right)}\Big]\mathcal{N}\Big\}}\ ,
    \end{aligned}
\end{equation}
\begin{equation}
    \begin{aligned}
         f_4 = \int_{-\infty}^{\infty}\textrm{d}X\, \frac{\frac{\mathrm{Pe}}{\sqrt{\epsilon}H^4} \Big\{ \frac{6}{H^3} \mathcal{N}\Big\}(\Xi_{\mathrm{p}}+\Xi_{\mathrm{w}})\Big[ H -\frac{2}{K} \tanh{\Big( \frac{KH}{2}\Big)}\Big]}{1+\frac{\textrm{Pe}}{\sqrt{\epsilon}H}\Big\{ \frac{\textrm{Ha}}{K^2}\mathcal{M}-\frac{6}{H^3}\frac{\textrm{Ha}}{K^2}(\Xi_{\textrm{p}}+\Xi_{\textrm{w}})\Big[ \frac{H}{2}-\frac{1}{K}\tanh{\left(\frac{KH}{2}\right)}\Big]\mathcal{N}\Big\}}\ ,
    \end{aligned}
\end{equation}
from which we can determine:
\begin{align}
   \mathcal{P}=\frac{-1}{\Big(\frac{9 \pi}{2 \Delta^{5/2}}+6 \frac{\mathrm{Ha}}{K^2}\frac{\mathrm{Pe}}{\sqrt{\epsilon}} f_4\Big)} \Big[ \frac{9 \pi X_{\textrm{G}} \dot{\Delta}}{2 \Delta^{5/2}}-\frac{3 \pi \dot{X}_{\textrm{G}}}{2 \Delta^{3/2}}+\frac{3 \pi (\dot{X}_{\textrm{G}}+\dot{\Theta})}{\Delta^{3/2}}+6\frac{\mathrm{Ha}}{K^2}\frac{\mathrm{Pe}}{\sqrt{\epsilon}} (\dot{X}_{\textrm{G}} f_1+\dot{\Theta} f_2 + \dot{\Delta} f_3 )\Big]. 
\end{align}
\end{widetext}

\section{Decomposition of potential}
\label{AppendixB}
\setcounter{equation}{0}
\renewcommand{\theequation}{\thesection\arabic{equation}}
In electrokinetic systems, it is convenient to decompose the total electric potential into an equilibrium component and a flow-induced potential, as:
\begin{equation}
\phi=\phi_0(x,z,t)+\phi_1(x,t).
\end{equation}
Here, $\phi_0$ denotes the equilibrium electric double-layer potential established by the charged surfaces in the absence of fluid motion. When the fluid is set into motion, the ions are advected by the flow. To maintain charge conservation, an additional streaming potential is generated. This contribution is represented by the perturbation potential $\phi_1$. Assuming that the flow-induced perturbation potential remains small compared with the equilibrium electric-double-layer potential, the governing electrokinetic equations can be linearised about the equilibrium state. The total electric potential can then be written as a superposition. This decomposition is widely employed in streaming-potential analyses because it separates the equilibrium electrostatic problem from the non-equilibrium transport processes generated by fluid motion.

\section{Forces and torques per unit length} \label{AppendixC}
\setcounter{equation}{0}
\renewcommand{\theequation}{\thesection\arabic{equation}}

The total stress tensor $\boldsymbol{\sigma}$ is given in dimensional units by:\\
\begin{widetext}
\begin{align}
   \boldsymbol{\sigma}=- p \, \mathbf{I} + \eta \left[\nabla \mathbf{u} + (\nabla \mathbf{u})^{\textrm{T}} \right]+ \varepsilon \left( \mathbf{E}\mathbf{E}-\frac{1}{2} E^2 \mathbf{I}\right)\ ,
\end{align}
where $\mathbf{E}$ is the electric field, $\mathbf{u}=(u,v)$ is the fluid velocity vector field, $\nabla$ is the nabla operator, $\mathbf{I}$ is the identity tensor, and where the superscript $\textrm{T}$ indicates transposition.
We can then write the traction vector exerted on the cylinder's surface as $\boldsymbol{\sigma}|_{z=h} \cdot \mathbf{n}$, where $\mathbf{n}$ is the normal vector to the cylinder's surface. Then, the components $k_x$ and $k_z$ of this traction vector, along the longitudinal and normal directions respectively, are given in dimensional units by:

\begin{align}
     k_x = -p|_{z=h} \ n_x + \eta \Big[\Big(\frac{\partial u}{\partial z} + \frac{\partial v}{\partial x} \Big) \ n_z + 2 \frac{\partial u}{\partial x} \ n_x\Big]_{z=h}+ \frac{\varepsilon}{2} \Big[2E_x E_z \ n_z + (E_x^2 -E_z^2) \ n_x\Big]_{z=h}\ ,\\
      k_z = -p|_{z=h} \ n_z  + \eta \Big[ \Big(\frac{\partial u}{\partial z} + \frac{\partial v}{\partial x} \Big) \ n_x + 2 \frac{\partial v}{\partial z} \ n_z\Big]_{z=h} + \frac{\varepsilon}{2} \Big[2E_x E_z \ n_x + (E_z^2 -E_x^2) \ n_z\Big]_{z=h}\ , 
\end{align}
where the normal vector to the surface of the cylinder can be approximated as
$\mathbf{n}=(n_{x}, n_{z}) \simeq [\sqrt{2\epsilon}(X-X_{\textrm{G}}),-1]$ in the lubrication approximation.
Invoking Eq.~(\ref{adim}), one also gets:
\begin{equation}
\begin{aligned}
    k_x = \frac{2\eta c}{a \epsilon} \Big\{-(X-X_{\textrm{G}})P +\Big[ \epsilon(X-X_{\textrm{G}})\frac{\partial U}{\partial X}-\frac{1}{2}\frac{\partial U}{\partial Z} - \frac{\epsilon}{2} \frac{\partial V}{\partial X}\Big]- \\
    \frac{1}{2}\frac{\mathrm{Ha}}{K^2} \Big(\Big[\Big( \frac{\partial \Phi_0}{\partial Z}\Big)^2 -\frac{\epsilon}{2} \Big(\frac{\partial \Phi_0}{\partial X}+\frac{\partial \Phi_1}{\partial X}\Big)^2\Big](X-X_{\textrm{G}})+\Big(\frac{\partial \Phi_0}{\partial X}+\frac{\partial \Phi_1}{\partial X}\Big)\frac{\partial \Phi_0}{\partial Z} \Big)\Big\}_{Z=H}\ , \label{kx}
\end{aligned}
\end{equation}
and:
\begin{equation}
\begin{aligned}
    k_z = \frac{\sqrt{2}\eta c}{a\epsilon^{3/2}}\Big\{P + \Big(\frac{1}{2}\frac{\partial U}{\partial Z} +\frac{\epsilon}{4} \frac{\partial V}{\partial X}\Big)[2\epsilon(X-X_{\textrm{G}})]-\epsilon \frac{\partial V}{\partial Z} +\\
    \frac{1}{2}\frac{\mathrm{Ha}}{K^2} \Big[2\epsilon(X-X_{\textrm{G}})\Big(\frac{\partial \Phi_0}{\partial X}+\frac{\partial \Phi_1}{\partial X}\Big)\frac{\partial \Phi_0}{\partial Z} - \Big( \frac{\partial \Phi_0}{\partial Z}\Big)^2 +\frac{\epsilon}{2} \Big(\frac{\partial \Phi_0}{\partial X}+\frac{\partial \Phi_1}{\partial X}\Big)^2\Big]\Big\}_{Z=H}\ . \label{kz}
\end{aligned}
\end{equation}
Altogether, the longitudinal force per unit length $f_x $, the normal force per unit length $f_z$, and the torque per unit length $\Gamma$, exerted on the cylinder, can finally be obtained in dimensional units from Eqs.~(\ref{kx}, \ref{kz}), as:
\begin{equation}
    \begin{aligned}
                 f_x &= \int_{-\infty}^{\infty} \textrm{d}x\,k_x \\
                  f_z &= \int_{-\infty}^{\infty} \textrm{d}x\,k_z \\
        \Gamma &= a\int_{-\infty}^{\infty}\textrm{d}x\, (n_x k_z - n_z k_x) \ .
    \end{aligned}
\end{equation}
\end{widetext}
Non-dimensionalizing these expressions and invoking linear and angular momentum conservation finally allows us to get the leading-order dimensionless equations of motion at stake, \textit{i.e.} Eqs.~(\ref{eqperp}-\ref{eqrot}). Note that we kept the $O(\epsilon)$ electrokinetic stress terms, because we have in mind the $\textrm{Pe}/\sqrt{\epsilon}$ factor in the relation between $\Phi_1$ and $\Phi_0$, as discussed after Eq.~(\ref{adim}). 
\bibliographystyle{aipnum4-2}
\bibliography{Chatterjee2026}

\begin{thebibliography}{49}%
\makeatletter
\providecommand \@ifxundefined [1]{%
 \@ifx{#1\undefined}
}%
\providecommand \@ifnum [1]{%
 \ifnum #1\expandafter \@firstoftwo
 \else \expandafter \@secondoftwo
 \fi
}%
\providecommand \@ifx [1]{%
 \ifx #1\expandafter \@firstoftwo
 \else \expandafter \@secondoftwo
 \fi
}%
\providecommand \natexlab [1]{#1}%
\providecommand \enquote  [1]{``#1''}%
\providecommand \bibnamefont  [1]{#1}%
\providecommand \bibfnamefont [1]{#1}%
\providecommand \citenamefont [1]{#1}%
\providecommand \href@noop [0]{\@secondoftwo}%
\providecommand \href [0]{\begingroup \@sanitize@url \@href}%
\providecommand \@href[1]{\@@startlink{#1}\@@href}%
\providecommand \@@href[1]{\endgroup#1\@@endlink}%
\providecommand \@sanitize@url [0]{\catcode `\\12\catcode `\$12\catcode
  `\&12\catcode `\#12\catcode `\^12\catcode `\_12\catcode `\%12\relax}%
\providecommand \@@startlink[1]{}%
\providecommand \@@endlink[0]{}%
\providecommand \url  [0]{\begingroup\@sanitize@url \@url }%
\providecommand \@url [1]{\endgroup\@href {#1}{\urlprefix }}%
\providecommand \urlprefix  [0]{URL }%
\providecommand \Eprint [0]{\href }%
\providecommand \doibase [0]{https://doi.org/}%
\providecommand \selectlanguage [0]{\@gobble}%
\providecommand \bibinfo  [0]{\@secondoftwo}%
\providecommand \bibfield  [0]{\@secondoftwo}%
\providecommand \translation [1]{[#1]}%
\providecommand \BibitemOpen [0]{}%
\providecommand \bibitemStop [0]{}%
\providecommand \bibitemNoStop [0]{.\EOS\space}%
\providecommand \EOS [0]{\spacefactor3000\relax}%
\providecommand \BibitemShut  [1]{\csname bibitem#1\endcsname}%
\let\auto@bib@innerbib\@empty
\bibitem [{\citenamefont {Leal}(2007)}]{Leal2007}%
  \BibitemOpen
  \bibfield  {author} {\bibinfo {author} {\bibfnamefont {L.~G.}\ \bibnamefont
  {Leal}},\ }\href@noop {} {\emph {\bibinfo {title} {Advanced Transport
  Phenomena: Fluid Mechanics and Convective Transport Processes}}}\ (\bibinfo
  {publisher} {Cambridge University Press},\ \bibinfo {address} {Cambridge},\
  \bibinfo {year} {2007})\BibitemShut {NoStop}%
\bibitem [{\citenamefont {Happel}\ and\ \citenamefont
  {Brenner}(1983)}]{HappelBrenner1983}%
  \BibitemOpen
  \bibfield  {author} {\bibinfo {author} {\bibfnamefont {J.}~\bibnamefont
  {Happel}}\ and\ \bibinfo {author} {\bibfnamefont {H.}~\bibnamefont
  {Brenner}},\ }\href@noop {} {\emph {\bibinfo {title} {Low Reynolds Number
  Hydrodynamics}}}\ (\bibinfo  {publisher} {Martinus Nijhoff},\ \bibinfo
  {address} {The Hague},\ \bibinfo {year} {1983})\BibitemShut {NoStop}%
\bibitem [{\citenamefont {Bocquet}\ and\ \citenamefont
  {Charlaix}(2010)}]{bocquet2010nanofluidics}%
  \BibitemOpen
  \bibfield  {author} {\bibinfo {author} {\bibfnamefont {L.}~\bibnamefont
  {Bocquet}}\ and\ \bibinfo {author} {\bibfnamefont {E.}~\bibnamefont
  {Charlaix}},\ }\href@noop {} {\bibfield  {journal} {\bibinfo  {journal}
  {Chemical Society Reviews}\ }\textbf {\bibinfo {volume} {39}},\ \bibinfo
  {pages} {1073} (\bibinfo {year} {2010})}\BibitemShut {NoStop}%
\bibitem [{\citenamefont {Skotheim}\ and\ \citenamefont
  {Mahadevan}(2005)}]{Skotheim2005}%
  \BibitemOpen
  \bibfield  {author} {\bibinfo {author} {\bibfnamefont {J.~M.}\ \bibnamefont
  {Skotheim}}\ and\ \bibinfo {author} {\bibfnamefont {L.}~\bibnamefont
  {Mahadevan}},\ }\href@noop {} {\bibfield  {journal} {\bibinfo  {journal}
  {Physics of Fluids}\ }\textbf {\bibinfo {volume} {17}},\ \bibinfo {pages}
  {092101} (\bibinfo {year} {2005})}\BibitemShut {NoStop}%
\bibitem [{\citenamefont {Bureau}, \citenamefont {Coupier},\ and\ \citenamefont
  {Salez}(2023)}]{Bureau2023}%
  \BibitemOpen
  \bibfield  {author} {\bibinfo {author} {\bibfnamefont {L.}~\bibnamefont
  {Bureau}}, \bibinfo {author} {\bibfnamefont {G.}~\bibnamefont {Coupier}},\
  and\ \bibinfo {author} {\bibfnamefont {T.}~\bibnamefont {Salez}},\
  }\href@noop {} {\bibfield  {journal} {\bibinfo  {journal} {European Physical
  Journal E}\ }\textbf {\bibinfo {volume} {46}},\ \bibinfo {pages} {111}
  (\bibinfo {year} {2023})}\BibitemShut {NoStop}%
\bibitem [{\citenamefont {Rallabandi}(2024)}]{Rallabandi2024}%
  \BibitemOpen
  \bibfield  {author} {\bibinfo {author} {\bibfnamefont {B.}~\bibnamefont
  {Rallabandi}},\ }\href@noop {} {\bibfield  {journal} {\bibinfo  {journal}
  {Annual Review of Fluid Mechanics}\ }\textbf {\bibinfo {volume} {56}},\
  \bibinfo {pages} {491} (\bibinfo {year} {2024})}\BibitemShut {NoStop}%
\bibitem [{\citenamefont {Saintyves}\ \emph {et~al.}(2016)\citenamefont
  {Saintyves}, \citenamefont {Jules}, \citenamefont {Salez},\ and\
  \citenamefont {Mahadevan}}]{Saintyves2016}%
  \BibitemOpen
  \bibfield  {author} {\bibinfo {author} {\bibfnamefont {B.}~\bibnamefont
  {Saintyves}}, \bibinfo {author} {\bibfnamefont {T.}~\bibnamefont {Jules}},
  \bibinfo {author} {\bibfnamefont {T.}~\bibnamefont {Salez}},\ and\ \bibinfo
  {author} {\bibfnamefont {L.}~\bibnamefont {Mahadevan}},\ }\href@noop {}
  {\bibfield  {journal} {\bibinfo  {journal} {Proceedings of the National
  Academy of Sciences}\ }\textbf {\bibinfo {volume} {113}},\ \bibinfo {pages}
  {5847} (\bibinfo {year} {2016})}\BibitemShut {NoStop}%
\bibitem [{\citenamefont {Davies-Strickleton}\ \emph
  {et~al.}(2018)\citenamefont {Davies-Strickleton}, \citenamefont
  {D{\'e}barre}, \citenamefont {El~Amri}, \citenamefont {Verdier},
  \citenamefont {Richter},\ and\ \citenamefont
  {Bureau}}]{davies2018elastohydrodynamic}%
  \BibitemOpen
  \bibfield  {author} {\bibinfo {author} {\bibfnamefont {H.}~\bibnamefont
  {Davies-Strickleton}}, \bibinfo {author} {\bibfnamefont {D.}~\bibnamefont
  {D{\'e}barre}}, \bibinfo {author} {\bibfnamefont {N.}~\bibnamefont
  {El~Amri}}, \bibinfo {author} {\bibfnamefont {C.}~\bibnamefont {Verdier}},
  \bibinfo {author} {\bibfnamefont {R.~P.}\ \bibnamefont {Richter}},\ and\
  \bibinfo {author} {\bibfnamefont {L.}~\bibnamefont {Bureau}},\ }\href@noop {}
  {\bibfield  {journal} {\bibinfo  {journal} {Physical review letters}\
  }\textbf {\bibinfo {volume} {120}},\ \bibinfo {pages} {198001} (\bibinfo
  {year} {2018})}\BibitemShut {NoStop}%
\bibitem [{\citenamefont {Rallabandi}\ \emph {et~al.}(2018)\citenamefont
  {Rallabandi}, \citenamefont {Oppenheimer}, \citenamefont {Ben~Zion},\ and\
  \citenamefont {Stone}}]{rallabandi2018membrane}%
  \BibitemOpen
  \bibfield  {author} {\bibinfo {author} {\bibfnamefont {B.}~\bibnamefont
  {Rallabandi}}, \bibinfo {author} {\bibfnamefont {N.}~\bibnamefont
  {Oppenheimer}}, \bibinfo {author} {\bibfnamefont {M.~Y.}\ \bibnamefont
  {Ben~Zion}},\ and\ \bibinfo {author} {\bibfnamefont {H.~A.}\ \bibnamefont
  {Stone}},\ }\href@noop {} {\bibfield  {journal} {\bibinfo  {journal} {Nature
  Physics}\ }\textbf {\bibinfo {volume} {14}},\ \bibinfo {pages} {1211}
  (\bibinfo {year} {2018})}\BibitemShut {NoStop}%
\bibitem [{\citenamefont {Vialar}\ \emph {et~al.}(2019)\citenamefont {Vialar},
  \citenamefont {Merzeau}, \citenamefont {Giasson},\ and\ \citenamefont
  {Drummond}}]{vialar2019compliant}%
  \BibitemOpen
  \bibfield  {author} {\bibinfo {author} {\bibfnamefont {P.}~\bibnamefont
  {Vialar}}, \bibinfo {author} {\bibfnamefont {P.}~\bibnamefont {Merzeau}},
  \bibinfo {author} {\bibfnamefont {S.}~\bibnamefont {Giasson}},\ and\ \bibinfo
  {author} {\bibfnamefont {C.}~\bibnamefont {Drummond}},\ }\href@noop {}
  {\bibfield  {journal} {\bibinfo  {journal} {Langmuir}\ }\textbf {\bibinfo
  {volume} {35}},\ \bibinfo {pages} {15605} (\bibinfo {year}
  {2019})}\BibitemShut {NoStop}%
\bibitem [{\citenamefont {Zhang}\ \emph {et~al.}(2020)\citenamefont {Zhang},
  \citenamefont {Bertin}, \citenamefont {Arshad}, \citenamefont {Rapha{\"e}l},
  \citenamefont {Salez},\ and\ \citenamefont {Maali}}]{zhang2020direct}%
  \BibitemOpen
  \bibfield  {author} {\bibinfo {author} {\bibfnamefont {Z.}~\bibnamefont
  {Zhang}}, \bibinfo {author} {\bibfnamefont {V.}~\bibnamefont {Bertin}},
  \bibinfo {author} {\bibfnamefont {M.}~\bibnamefont {Arshad}}, \bibinfo
  {author} {\bibfnamefont {E.}~\bibnamefont {Rapha{\"e}l}}, \bibinfo {author}
  {\bibfnamefont {T.}~\bibnamefont {Salez}},\ and\ \bibinfo {author}
  {\bibfnamefont {A.}~\bibnamefont {Maali}},\ }\href@noop {} {\bibfield
  {journal} {\bibinfo  {journal} {Physical review letters}\ }\textbf {\bibinfo
  {volume} {124}},\ \bibinfo {pages} {054502} (\bibinfo {year}
  {2020})}\BibitemShut {NoStop}%
\bibitem [{\citenamefont {Fares}\ \emph {et~al.}(2024)\citenamefont {Fares},
  \citenamefont {Lavaud}, \citenamefont {Zhang}, \citenamefont {Jha},
  \citenamefont {Amarouchene},\ and\ \citenamefont
  {Salez}}]{fares2024observation}%
  \BibitemOpen
  \bibfield  {author} {\bibinfo {author} {\bibfnamefont {N.}~\bibnamefont
  {Fares}}, \bibinfo {author} {\bibfnamefont {M.}~\bibnamefont {Lavaud}},
  \bibinfo {author} {\bibfnamefont {Z.}~\bibnamefont {Zhang}}, \bibinfo
  {author} {\bibfnamefont {A.}~\bibnamefont {Jha}}, \bibinfo {author}
  {\bibfnamefont {Y.}~\bibnamefont {Amarouchene}},\ and\ \bibinfo {author}
  {\bibfnamefont {T.}~\bibnamefont {Salez}},\ }\href@noop {} {\bibfield
  {journal} {\bibinfo  {journal} {Proceedings of the National Academy of
  Sciences}\ }\textbf {\bibinfo {volume} {121}},\ \bibinfo {pages}
  {e2411956121} (\bibinfo {year} {2024})}\BibitemShut {NoStop}%
\bibitem [{\citenamefont {Karan}, \citenamefont {Yariv},\ and\ \citenamefont
  {Rallabandi}(2020)}]{Karan2020}%
  \BibitemOpen
  \bibfield  {author} {\bibinfo {author} {\bibfnamefont {P.}~\bibnamefont
  {Karan}}, \bibinfo {author} {\bibfnamefont {E.}~\bibnamefont {Yariv}},\ and\
  \bibinfo {author} {\bibfnamefont {B.}~\bibnamefont {Rallabandi}},\
  }\href@noop {} {\bibfield  {journal} {\bibinfo  {journal} {Physical Review
  Fluids}\ }\textbf {\bibinfo {volume} {5}},\ \bibinfo {pages} {082001}
  (\bibinfo {year} {2020})}\BibitemShut {NoStop}%
\bibitem [{\citenamefont {Rinehart}\ \emph {et~al.}(2020)\citenamefont
  {Rinehart}, \citenamefont {L{\=a}cis}, \citenamefont {Salez},\ and\
  \citenamefont {Bagheri}}]{rinehart2020lift}%
  \BibitemOpen
  \bibfield  {author} {\bibinfo {author} {\bibfnamefont {A.}~\bibnamefont
  {Rinehart}}, \bibinfo {author} {\bibfnamefont {U.}~\bibnamefont {L{\=a}cis}},
  \bibinfo {author} {\bibfnamefont {T.}~\bibnamefont {Salez}},\ and\ \bibinfo
  {author} {\bibfnamefont {S.}~\bibnamefont {Bagheri}},\ }\href@noop {}
  {\bibfield  {journal} {\bibinfo  {journal} {Physical Review Fluids}\ }\textbf
  {\bibinfo {volume} {5}},\ \bibinfo {pages} {082001} (\bibinfo {year}
  {2020})}\BibitemShut {NoStop}%
\bibitem [{\citenamefont {Pandey}\ \emph {et~al.}(2016)\citenamefont {Pandey},
  \citenamefont {Karpitschka}, \citenamefont {Venner},\ and\ \citenamefont
  {Snoeijer}}]{Pandey2016}%
  \BibitemOpen
  \bibfield  {author} {\bibinfo {author} {\bibfnamefont {A.}~\bibnamefont
  {Pandey}}, \bibinfo {author} {\bibfnamefont {S.}~\bibnamefont {Karpitschka}},
  \bibinfo {author} {\bibfnamefont {C.~H.}\ \bibnamefont {Venner}},\ and\
  \bibinfo {author} {\bibfnamefont {J.~H.}\ \bibnamefont {Snoeijer}},\
  }\href@noop {} {\bibfield  {journal} {\bibinfo  {journal} {Journal of Fluid
  Mechanics}\ }\textbf {\bibinfo {volume} {799}},\ \bibinfo {pages} {433}
  (\bibinfo {year} {2016})}\BibitemShut {NoStop}%
\bibitem [{\citenamefont {Bharti}\ \emph {et~al.}(2024)\citenamefont {Bharti}
  \emph {et~al.}}]{Bharti2024}%
  \BibitemOpen
  \bibfield  {author} {\bibinfo {author} {\bibnamefont {Bharti}} \emph
  {et~al.},\ }\href@noop {} {\bibfield  {journal} {\bibinfo  {journal}
  {Physical Review Research}\ }\textbf {\bibinfo {volume} {6}},\ \bibinfo
  {pages} {043060} (\bibinfo {year} {2024})}\BibitemShut {NoStop}%
\bibitem [{\citenamefont {Oratis}\ \emph {et~al.}(2025)\citenamefont {Oratis},
  \citenamefont {Bertin}, \citenamefont {Kansal},\ and\ \citenamefont
  {Snoeijer}}]{oratis2025viscoelastic}%
  \BibitemOpen
  \bibfield  {author} {\bibinfo {author} {\bibfnamefont {A.~T.}\ \bibnamefont
  {Oratis}}, \bibinfo {author} {\bibfnamefont {V.}~\bibnamefont {Bertin}},
  \bibinfo {author} {\bibfnamefont {M.}~\bibnamefont {Kansal}},\ and\ \bibinfo
  {author} {\bibfnamefont {J.~H.}\ \bibnamefont {Snoeijer}},\ }in\ \href@noop
  {} {\emph {\bibinfo {booktitle} {Interfacial Flows?The Power and Beauty of
  Asymptotic Methods}}}\ (\bibinfo  {publisher} {Springer},\ \bibinfo {year}
  {2025})\ pp.\ \bibinfo {pages} {175--195}\BibitemShut {NoStop}%
\bibitem [{\citenamefont {Lou}\ \emph {et~al.}(2022)\citenamefont {Lou},
  \citenamefont {Yang}, \citenamefont {Ding}, \citenamefont {Liu},
  \citenamefont {Chen}, \citenamefont {Zhou}, \citenamefont {Ye}, \citenamefont
  {Podgornik},\ and\ \citenamefont {Yang}}]{lou2022odd}%
  \BibitemOpen
  \bibfield  {author} {\bibinfo {author} {\bibfnamefont {X.}~\bibnamefont
  {Lou}}, \bibinfo {author} {\bibfnamefont {Q.}~\bibnamefont {Yang}}, \bibinfo
  {author} {\bibfnamefont {Y.}~\bibnamefont {Ding}}, \bibinfo {author}
  {\bibfnamefont {P.}~\bibnamefont {Liu}}, \bibinfo {author} {\bibfnamefont
  {K.}~\bibnamefont {Chen}}, \bibinfo {author} {\bibfnamefont {X.}~\bibnamefont
  {Zhou}}, \bibinfo {author} {\bibfnamefont {F.}~\bibnamefont {Ye}}, \bibinfo
  {author} {\bibfnamefont {R.}~\bibnamefont {Podgornik}},\ and\ \bibinfo
  {author} {\bibfnamefont {M.}~\bibnamefont {Yang}},\ }\href@noop {} {\bibfield
   {journal} {\bibinfo  {journal} {Proceedings of the National Academy of
  Sciences}\ }\textbf {\bibinfo {volume} {119}},\ \bibinfo {pages}
  {e2201279119} (\bibinfo {year} {2022})}\BibitemShut {NoStop}%
\bibitem [{\citenamefont {Small}(1974)}]{Small1974}%
  \BibitemOpen
  \bibfield  {author} {\bibinfo {author} {\bibfnamefont {H.}~\bibnamefont
  {Small}},\ }\href@noop {} {\bibfield  {journal} {\bibinfo  {journal} {Journal
  of Colloid and Interface Science}\ }\textbf {\bibinfo {volume} {48}},\
  \bibinfo {pages} {147} (\bibinfo {year} {1974})}\BibitemShut {NoStop}%
\bibitem [{\citenamefont {Giddings}(1991)}]{Giddings1993}%
  \BibitemOpen
  \bibfield  {author} {\bibinfo {author} {\bibfnamefont {J.~C.}\ \bibnamefont
  {Giddings}},\ }\href@noop {} {\emph {\bibinfo {title} {Unified Separation
  Science}}}\ (\bibinfo  {publisher} {Wiley},\ \bibinfo {address} {New York},\
  \bibinfo {year} {1991})\BibitemShut {NoStop}%
\bibitem [{\citenamefont {Li}\ \emph {et~al.}(2014)\citenamefont {Li},
  \citenamefont {Li}, \citenamefont {Zhang}, \citenamefont {Alici},\ and\
  \citenamefont {Wen}}]{Li2014}%
  \BibitemOpen
  \bibfield  {author} {\bibinfo {author} {\bibfnamefont {M.}~\bibnamefont
  {Li}}, \bibinfo {author} {\bibfnamefont {W.~H.}\ \bibnamefont {Li}}, \bibinfo
  {author} {\bibfnamefont {J.}~\bibnamefont {Zhang}}, \bibinfo {author}
  {\bibfnamefont {G.}~\bibnamefont {Alici}},\ and\ \bibinfo {author}
  {\bibfnamefont {W.}~\bibnamefont {Wen}},\ }\href@noop {} {\bibfield
  {journal} {\bibinfo  {journal} {Journal of Physics D: Applied Physics}\
  }\textbf {\bibinfo {volume} {47}},\ \bibinfo {pages} {063001} (\bibinfo
  {year} {2014})}\BibitemShut {NoStop}%
\bibitem [{\citenamefont {Xuan}(2019)}]{Xuan2019}%
  \BibitemOpen
  \bibfield  {author} {\bibinfo {author} {\bibfnamefont {X.}~\bibnamefont
  {Xuan}},\ }\href@noop {} {\bibfield  {journal} {\bibinfo  {journal}
  {Electrophoresis}\ }\textbf {\bibinfo {volume} {40}},\ \bibinfo {pages}
  {2484} (\bibinfo {year} {2019})}\BibitemShut {NoStop}%
\bibitem [{\citenamefont {Zhang}\ and\ \citenamefont
  {Umehara}(1998)}]{ZhangUmehara1998}%
  \BibitemOpen
  \bibfield  {author} {\bibinfo {author} {\bibfnamefont {B.}~\bibnamefont
  {Zhang}}\ and\ \bibinfo {author} {\bibfnamefont {N.}~\bibnamefont
  {Umehara}},\ }\href@noop {} {\bibfield  {journal} {\bibinfo  {journal} {JSME
  International Journal Series C}\ }\textbf {\bibinfo {volume} {41}},\ \bibinfo
  {pages} {285} (\bibinfo {year} {1998})}\BibitemShut {NoStop}%
\bibitem [{\citenamefont {Wong}, \citenamefont {Huang},\ and\ \citenamefont
  {Meng}(2003)}]{WongETal2003}%
  \BibitemOpen
  \bibfield  {author} {\bibinfo {author} {\bibfnamefont {P.~L.}\ \bibnamefont
  {Wong}}, \bibinfo {author} {\bibfnamefont {P.}~\bibnamefont {Huang}},\ and\
  \bibinfo {author} {\bibfnamefont {Y.}~\bibnamefont {Meng}},\ }\href@noop {}
  {\bibfield  {journal} {\bibinfo  {journal} {Tribology Letters}\ }\textbf
  {\bibinfo {volume} {14}},\ \bibinfo {pages} {197} (\bibinfo {year}
  {2003})}\BibitemShut {NoStop}%
\bibitem [{\citenamefont {Elton}(1948)}]{Elton1948}%
  \BibitemOpen
  \bibfield  {author} {\bibinfo {author} {\bibfnamefont {G.~A.~H.}\
  \bibnamefont {Elton}},\ }\href@noop {} {\bibfield  {journal} {\bibinfo
  {journal} {Proceedings of the Royal Society A}\ }\textbf {\bibinfo {volume}
  {194}},\ \bibinfo {pages} {296} (\bibinfo {year} {1948})}\BibitemShut
  {NoStop}%
\bibitem [{\citenamefont {Rice}\ and\ \citenamefont
  {Whitehead}(1965)}]{RiceWhitehead1965}%
  \BibitemOpen
  \bibfield  {author} {\bibinfo {author} {\bibfnamefont {C.~L.}\ \bibnamefont
  {Rice}}\ and\ \bibinfo {author} {\bibfnamefont {R.}~\bibnamefont
  {Whitehead}},\ }\href@noop {} {\bibfield  {journal} {\bibinfo  {journal}
  {Journal of Physical Chemistry}\ }\textbf {\bibinfo {volume} {69}},\ \bibinfo
  {pages} {4017} (\bibinfo {year} {1965})}\BibitemShut {NoStop}%
\bibitem [{\citenamefont {Levine}\ \emph {et~al.}(1975)\citenamefont {Levine},
  \citenamefont {Marriott}, \citenamefont {Neale},\ and\ \citenamefont
  {Epstein}}]{Levine1975}%
  \BibitemOpen
  \bibfield  {author} {\bibinfo {author} {\bibfnamefont {S.}~\bibnamefont
  {Levine}}, \bibinfo {author} {\bibfnamefont {J.~R.}\ \bibnamefont
  {Marriott}}, \bibinfo {author} {\bibfnamefont {G.}~\bibnamefont {Neale}},\
  and\ \bibinfo {author} {\bibfnamefont {N.}~\bibnamefont {Epstein}},\
  }\href@noop {} {\bibfield  {journal} {\bibinfo  {journal} {Journal of Colloid
  and Interface Science}\ }\textbf {\bibinfo {volume} {52}},\ \bibinfo {pages}
  {136} (\bibinfo {year} {1975})}\BibitemShut {NoStop}%
\bibitem [{\citenamefont {Hunter}(1981)}]{Hunter1981}%
  \BibitemOpen
  \bibfield  {author} {\bibinfo {author} {\bibfnamefont {R.~J.}\ \bibnamefont
  {Hunter}},\ }\href@noop {} {\emph {\bibinfo {title} {Zeta Potential in
  Colloid Science: Principles and Applications}}}\ (\bibinfo  {publisher}
  {Academic Press},\ \bibinfo {address} {New York},\ \bibinfo {year}
  {1981})\BibitemShut {NoStop}%
\bibitem [{\citenamefont {Russel}, \citenamefont {Saville},\ and\ \citenamefont
  {Schowalter}(1989)}]{RusselSavilleSchowalter1989}%
  \BibitemOpen
  \bibfield  {author} {\bibinfo {author} {\bibfnamefont {W.~B.}\ \bibnamefont
  {Russel}}, \bibinfo {author} {\bibfnamefont {D.~A.}\ \bibnamefont
  {Saville}},\ and\ \bibinfo {author} {\bibfnamefont {W.~R.}\ \bibnamefont
  {Schowalter}},\ }\href@noop {} {\emph {\bibinfo {title} {Colloidal
  Dispersions}}}\ (\bibinfo  {publisher} {Cambridge University Press},\
  \bibinfo {address} {Cambridge},\ \bibinfo {year} {1989})\BibitemShut
  {NoStop}%
\bibitem [{\citenamefont {Li}\ and\ \citenamefont {Jin}(2008)}]{LiJin2008}%
  \BibitemOpen
  \bibfield  {author} {\bibinfo {author} {\bibfnamefont {W.-L.}\ \bibnamefont
  {Li}}\ and\ \bibinfo {author} {\bibfnamefont {Z.}~\bibnamefont {Jin}},\
  }\href@noop {} {\bibfield  {journal} {\bibinfo  {journal} {Proceedings of the
  Institution of Mechanical Engineers Part J}\ }\textbf {\bibinfo {volume}
  {222}},\ \bibinfo {pages} {201} (\bibinfo {year} {2008})}\BibitemShut
  {NoStop}%
\bibitem [{\citenamefont {Chakraborty}\ and\ \citenamefont
  {Chakraborty}(2011)}]{ChakrabortyChakraborty2011}%
  \BibitemOpen
  \bibfield  {author} {\bibinfo {author} {\bibfnamefont {J.}~\bibnamefont
  {Chakraborty}}\ and\ \bibinfo {author} {\bibfnamefont {S.}~\bibnamefont
  {Chakraborty}},\ }\href@noop {} {\bibfield  {journal} {\bibinfo  {journal}
  {Physics of Fluids}\ }\textbf {\bibinfo {volume} {23}},\ \bibinfo {pages}
  {082004} (\bibinfo {year} {2011})}\BibitemShut {NoStop}%
\bibitem [{\citenamefont {Cramail}(2024)}]{cramail2024forces}%
  \BibitemOpen
  \bibfield  {author} {\bibinfo {author} {\bibfnamefont {C.}~\bibnamefont
  {Cramail}},\ }\emph {\bibinfo {title} {Forces de surface dynamiques des
  solutions ioniques confin{\'e}es: mesures exp{\'e}rimentales et
  mod{\'e}lisation th{\'e}orique}},\ \href@noop {} {Ph.D. thesis},\ \bibinfo
  {school} {Universit{\'e} Grenoble Alpes} (\bibinfo {year} {2024})\BibitemShut
  {NoStop}%
\bibitem [{\citenamefont {Bike}\ and\ \citenamefont {Prieve}(1990)}]{Bike1990}%
  \BibitemOpen
  \bibfield  {author} {\bibinfo {author} {\bibfnamefont {S.~G.}\ \bibnamefont
  {Bike}}\ and\ \bibinfo {author} {\bibfnamefont {D.~C.}\ \bibnamefont
  {Prieve}},\ }\href@noop {} {\bibfield  {journal} {\bibinfo  {journal}
  {Journal of Colloid and Interface Science}\ }\textbf {\bibinfo {volume}
  {136}},\ \bibinfo {pages} {95} (\bibinfo {year} {1990})}\BibitemShut
  {NoStop}%
\bibitem [{\citenamefont {Warszy\'nski}\ and\ \citenamefont {van~de
  Ven}(2000)}]{Warszynski2000}%
  \BibitemOpen
  \bibfield  {author} {\bibinfo {author} {\bibfnamefont {P.}~\bibnamefont
  {Warszy\'nski}}\ and\ \bibinfo {author} {\bibfnamefont {T.~G.~M.}\
  \bibnamefont {van~de Ven}},\ }\href@noop {} {\bibfield  {journal} {\bibinfo
  {journal} {Journal of Colloid and Interface Science}\ }\textbf {\bibinfo
  {volume} {223}},\ \bibinfo {pages} {1} (\bibinfo {year} {2000})}\BibitemShut
  {NoStop}%
\bibitem [{\citenamefont {Tabatabaei}, \citenamefont {van~de Ven},\ and\
  \citenamefont {Rey}(2006)}]{Tabatabaei2006}%
  \BibitemOpen
  \bibfield  {author} {\bibinfo {author} {\bibfnamefont {S.~M.}\ \bibnamefont
  {Tabatabaei}}, \bibinfo {author} {\bibfnamefont {T.~G.~M.}\ \bibnamefont
  {van~de Ven}},\ and\ \bibinfo {author} {\bibfnamefont {A.~D.}\ \bibnamefont
  {Rey}},\ }\href@noop {} {\bibfield  {journal} {\bibinfo  {journal} {Journal
  of Colloid and Interface Science}\ }\textbf {\bibinfo {volume} {295}},\
  \bibinfo {pages} {504} (\bibinfo {year} {2006})}\BibitemShut {NoStop}%
\bibitem [{\citenamefont {Yariv}, \citenamefont {Schnitzer},\ and\
  \citenamefont {Frankel}(2011)}]{Yariv2011}%
  \BibitemOpen
  \bibfield  {author} {\bibinfo {author} {\bibfnamefont {E.}~\bibnamefont
  {Yariv}}, \bibinfo {author} {\bibfnamefont {O.}~\bibnamefont {Schnitzer}},\
  and\ \bibinfo {author} {\bibfnamefont {I.}~\bibnamefont {Frankel}},\
  }\href@noop {} {\bibfield  {journal} {\bibinfo  {journal} {Journal of Fluid
  Mechanics}\ }\textbf {\bibinfo {volume} {685}},\ \bibinfo {pages} {306}
  (\bibinfo {year} {2011})}\BibitemShut {NoStop}%
\bibitem [{\citenamefont {Schnitzer}, \citenamefont {Frankel},\ and\
  \citenamefont {Yariv}(2012)}]{Schnitzer2012}%
  \BibitemOpen
  \bibfield  {author} {\bibinfo {author} {\bibfnamefont {O.}~\bibnamefont
  {Schnitzer}}, \bibinfo {author} {\bibfnamefont {I.}~\bibnamefont {Frankel}},\
  and\ \bibinfo {author} {\bibfnamefont {E.}~\bibnamefont {Yariv}},\
  }\href@noop {} {\bibfield  {journal} {\bibinfo  {journal} {Journal of Fluid
  Mechanics}\ }\textbf {\bibinfo {volume} {704}},\ \bibinfo {pages} {109}
  (\bibinfo {year} {2012})}\BibitemShut {NoStop}%
\bibitem [{\citenamefont {Schnitzer}\ and\ \citenamefont
  {Yariv}(2016)}]{Schnitzer2016}%
  \BibitemOpen
  \bibfield  {author} {\bibinfo {author} {\bibfnamefont {O.}~\bibnamefont
  {Schnitzer}}\ and\ \bibinfo {author} {\bibfnamefont {E.}~\bibnamefont
  {Yariv}},\ }\href@noop {} {\bibfield  {journal} {\bibinfo  {journal} {Journal
  of Fluid Mechanics}\ }\textbf {\bibinfo {volume} {786}},\ \bibinfo {pages}
  {84} (\bibinfo {year} {2016})}\BibitemShut {NoStop}%
\bibitem [{\citenamefont {Liu}\ \emph {et~al.}(2018)\citenamefont {Liu} \emph
  {et~al.}}]{Liu2018}%
  \BibitemOpen
  \bibfield  {author} {\bibinfo {author} {\bibfnamefont {F.}~\bibnamefont
  {Liu}} \emph {et~al.},\ }\href@noop {} {\bibfield  {journal} {\bibinfo
  {journal} {Journal of Physical Chemistry B}\ }\textbf {\bibinfo {volume}
  {122}},\ \bibinfo {pages} {933} (\bibinfo {year} {2018})}\BibitemShut
  {NoStop}%
\bibitem [{\citenamefont {Zhao}\ \emph {et~al.}(2020)\citenamefont {Zhao} \emph
  {et~al.}}]{Zhao2020}%
  \BibitemOpen
  \bibfield  {author} {\bibinfo {author} {\bibfnamefont {C.}~\bibnamefont
  {Zhao}} \emph {et~al.},\ }\href@noop {} {\bibfield  {journal} {\bibinfo
  {journal} {Journal of Fluid Mechanics}\ }\textbf {\bibinfo {volume} {888}},\
  \bibinfo {pages} {A29} (\bibinfo {year} {2020})}\BibitemShut {NoStop}%
\bibitem [{\citenamefont {Rodriguez~Matus}\ \emph {et~al.}(2022)\citenamefont
  {Rodriguez~Matus} \emph {et~al.}}]{Matus2022}%
  \BibitemOpen
  \bibfield  {author} {\bibinfo {author} {\bibfnamefont {M.}~\bibnamefont
  {Rodriguez~Matus}} \emph {et~al.},\ }\href@noop {} {\bibfield  {journal}
  {\bibinfo  {journal} {Physical Review E}\ }\textbf {\bibinfo {volume}
  {105}},\ \bibinfo {pages} {064606} (\bibinfo {year} {2022})}\BibitemShut
  {NoStop}%
\bibitem [{\citenamefont {Chun}\ and\ \citenamefont
  {Ladd}(2004)}]{ChunLadd2004}%
  \BibitemOpen
  \bibfield  {author} {\bibinfo {author} {\bibfnamefont {B.}~\bibnamefont
  {Chun}}\ and\ \bibinfo {author} {\bibfnamefont {A.~J.~C.}\ \bibnamefont
  {Ladd}},\ }\href@noop {} {\bibfield  {journal} {\bibinfo  {journal} {Journal
  of Colloid and Interface Science}\ }\textbf {\bibinfo {volume} {274}},\
  \bibinfo {pages} {687} (\bibinfo {year} {2004})}\BibitemShut {NoStop}%
\bibitem [{\citenamefont {Alexander}\ and\ \citenamefont
  {Prieve}(1986)}]{AlexanderPrieve1986}%
  \BibitemOpen
  \bibfield  {author} {\bibinfo {author} {\bibfnamefont {B.~M.}\ \bibnamefont
  {Alexander}}\ and\ \bibinfo {author} {\bibfnamefont {D.~C.}\ \bibnamefont
  {Prieve}},\ }\href@noop {} {\bibfield  {journal} {\bibinfo  {journal}
  {Langmuir}\ }\textbf {\bibinfo {volume} {2}},\ \bibinfo {pages} {442}
  (\bibinfo {year} {1986})}\BibitemShut {NoStop}%
\bibitem [{\citenamefont {Bike}\ and\ \citenamefont
  {Prieve}(1995)}]{BikePrieve1995}%
  \BibitemOpen
  \bibfield  {author} {\bibinfo {author} {\bibfnamefont {S.~G.}\ \bibnamefont
  {Bike}}\ and\ \bibinfo {author} {\bibfnamefont {D.~C.}\ \bibnamefont
  {Prieve}},\ }\href@noop {} {\bibfield  {journal} {\bibinfo  {journal}
  {Journal of Colloid and Interface Science}\ }\textbf {\bibinfo {volume}
  {175}},\ \bibinfo {pages} {422} (\bibinfo {year} {1995})}\BibitemShut
  {NoStop}%
\bibitem [{\citenamefont {Bike}, \citenamefont {Lazarro},\ and\ \citenamefont
  {Prieve}(1995)}]{Bike2002}%
  \BibitemOpen
  \bibfield  {author} {\bibinfo {author} {\bibfnamefont {S.~G.}\ \bibnamefont
  {Bike}}, \bibinfo {author} {\bibfnamefont {L.}~\bibnamefont {Lazarro}},\ and\
  \bibinfo {author} {\bibfnamefont {D.~C.}\ \bibnamefont {Prieve}},\
  }\href@noop {} {\bibfield  {journal} {\bibinfo  {journal} {Journal of Colloid
  and Interface Science}\ }\textbf {\bibinfo {volume} {175}},\ \bibinfo {pages}
  {411} (\bibinfo {year} {1995})}\BibitemShut {NoStop}%
\bibitem [{\citenamefont {Salez}\ and\ \citenamefont
  {Mahadevan}(2015)}]{SalezMahadevan2015}%
  \BibitemOpen
  \bibfield  {author} {\bibinfo {author} {\bibfnamefont {T.}~\bibnamefont
  {Salez}}\ and\ \bibinfo {author} {\bibfnamefont {L.}~\bibnamefont
  {Mahadevan}},\ }\href@noop {} {\bibfield  {journal} {\bibinfo  {journal}
  {Journal of Fluid Mechanics}\ }\textbf {\bibinfo {volume} {779}},\ \bibinfo
  {pages} {181} (\bibinfo {year} {2015})}\BibitemShut {NoStop}%
\bibitem [{\citenamefont {Bertin}\ \emph {et~al.}(2022)\citenamefont {Bertin},
  \citenamefont {Amarouchene}, \citenamefont {Rapha\"el},\ and\ \citenamefont
  {Salez}}]{Bertin2022}%
  \BibitemOpen
  \bibfield  {author} {\bibinfo {author} {\bibfnamefont {V.}~\bibnamefont
  {Bertin}}, \bibinfo {author} {\bibfnamefont {Y.}~\bibnamefont {Amarouchene}},
  \bibinfo {author} {\bibfnamefont {E.}~\bibnamefont {Rapha\"el}},\ and\
  \bibinfo {author} {\bibfnamefont {T.}~\bibnamefont {Salez}},\ }\href@noop {}
  {\bibfield  {journal} {\bibinfo  {journal} {Journal of Fluid Mechanics}\
  }\textbf {\bibinfo {volume} {933}},\ \bibinfo {pages} {A23} (\bibinfo {year}
  {2022})}\BibitemShut {NoStop}%
\bibitem [{\citenamefont {Jha}, \citenamefont {Amarouchene},\ and\
  \citenamefont {Salez}(2024)}]{Jha2024}%
  \BibitemOpen
  \bibfield  {author} {\bibinfo {author} {\bibfnamefont {A.}~\bibnamefont
  {Jha}}, \bibinfo {author} {\bibfnamefont {Y.}~\bibnamefont {Amarouchene}},\
  and\ \bibinfo {author} {\bibfnamefont {T.}~\bibnamefont {Salez}},\
  }\href@noop {} {\bibfield  {journal} {\bibinfo  {journal} {Journal of Fluid
  Mechanics}\ }\textbf {\bibinfo {volume} {1001}},\ \bibinfo {pages} {A58}
  (\bibinfo {year} {2024})}\BibitemShut {NoStop}%
\bibitem [{\citenamefont {Ye}\ \emph {et~al.}(2025)\citenamefont {Ye},
  \citenamefont {Amarouchene}, \citenamefont {Sarfati}, \citenamefont {Dean},\
  and\ \citenamefont {Salez}}]{ye2025brownian}%
  \BibitemOpen
  \bibfield  {author} {\bibinfo {author} {\bibfnamefont {Y.}~\bibnamefont
  {Ye}}, \bibinfo {author} {\bibfnamefont {Y.}~\bibnamefont {Amarouchene}},
  \bibinfo {author} {\bibfnamefont {R.}~\bibnamefont {Sarfati}}, \bibinfo
  {author} {\bibfnamefont {D.~S.}\ \bibnamefont {Dean}},\ and\ \bibinfo
  {author} {\bibfnamefont {T.}~\bibnamefont {Salez}},\ }\href@noop {}
  {\bibfield  {journal} {\bibinfo  {journal} {Comptes Rendus. Physique}\
  }\textbf {\bibinfo {volume} {26}},\ \bibinfo {pages} {619} (\bibinfo {year}
  {2025})}\BibitemShut {NoStop}%
\end{thebibliography}%
\end{document}